\documentclass[journal]{IEEEtran}
\IEEEoverridecommandlockouts

\usepackage{cite}
\usepackage{amsmath,amssymb,amsthm}
\usepackage{graphicx}
\usepackage{subcaption}
\usepackage{url}
\usepackage{booktabs}
\usepackage{bm}
\usepackage{xcolor}
\usepackage{colortbl}
\usepackage{placeins}
\usepackage{hyperref}
\newcommand{\rev}[1]{#1}
\newcommand{\revt}[1]{#1}

\expandafter\def\expandafter\UrlBreaks\expandafter{\UrlBreaks\do\-}

\newtheorem{proposition}{Proposition}
\newtheorem{corollary}{Corollary}
\newtheorem{remark}{Remark}

\newcommand{\x}{\mathbf{x}}
\newcommand{\Rt}{\mathbb{R}}
\newcommand{\Tr}{\operatorname{Tr}}
\newcommand{\diag}{\operatorname{diag}}
\DeclareMathOperator*{\argmin}{arg\,min}
\newcommand{\E}{\mathbb{E}}
\newcommand{\KN}{K_{\mathrm{NLOS}}}
\definecolor{cmarkbg}{RGB}{209,240,209}
\definecolor{xmarkbg}{RGB}{250,214,214}
\definecolor{pmarkbg}{RGB}{250,240,200}
\newcommand{\cmark}{\cellcolor{cmarkbg}\checkmark}
\newcommand{\xmark}{\cellcolor{xmarkbg}\ensuremath{\times}}
\newcommand{\pmark}{\cellcolor{pmarkbg}\ensuremath{\sim}}

\begin{document}

\title{NLOS-Aware TDOA-Based UAV Localization: Cram\'er-Rao Bounds, Measurement Modeling, and Experimental Validation}

\author{Cole~Dickerson,~Saad~Masrur,~\"Ozg\"ur~\"Ozdemir,~and~Ismail~G\"uven\c{c}%
\thanks{C.~Dickerson, S.~Masrur, \"O.~\"Ozdemir, and I.~G\"uven\c{c} are with the
Department of Electrical and Computer Engineering, North Carolina State
University, Raleigh, NC 27606 USA (e-mail: \{jcdicker, smasrur, oozdemi,
iguvenc\}@ncsu.edu).}%
\thanks{This work was supported in part by the National Science Foundation (NSF)
under Grants CNS-2450593, CNS-2332834, and DGE-2137100. The material in this paper was presented in
part at the IEEE International Conference on Communications (ICC) Workshops, Montreal, QC, Canada,
Jun. 2025~\cite{dickerson2025icc}.}}

\markboth{IEEE Transactions on Aerospace and Electronic Systems}%
{Dickerson \MakeLowercase{\textit{et al.}}: Impact of Altitude, Bandwidth, and NLOS Bias on 3D UAV Localization}

\maketitle

\begin{abstract}
\rev{Passive time-difference-of-arrival} (TDOA) localization is well suited for tracking non-cooperative
\rev{unmanned aerial vehicles (UAVs)} that do not report their positions. However, its accuracy degrades under the mixed
line-of-sight (LOS) and non-line-of-sight (NLOS) propagation that is typical of low-altitude flight.
In this paper, we develop NLOS-aware Cram\'er--Rao lower bounds (CRLBs) and an empirically
calibrated measurement model for TDOA-based UAV localization, and we evaluate both with measurements
from five UAV flights on the Aerial Experimentation and Research Platform for Advanced Wireless
(AERPAW). The analysis yields an identifiability condition that relates the number of sensors to the
number of NLOS links, together with a bias-induced error floor that a wider bandwidth cannot remove.
The model parameters are estimated by matching simulated and measured error distributions.
Experimental results show that the error depends most strongly on the number of sensors in LOS,
while a higher altitude and a wider bandwidth lower the median error. Relocating a \rev{sensor that is always in NLOS} into the open flight area can lower the median horizontal bound by two orders of
magnitude if the relocated sensor is in LOS. The fitted model reproduces the median errors of the
\rev{2.5~MHz and 5~MHz} flights but underestimates the heaviest error tails. \rev{Finally, rejecting
outlier fixes lowers the error far more than Kalman or particle filtering, because the remaining
error is dominated by bias.}
\end{abstract}


\section{Introduction}
\IEEEPARstart{T}{he} detection and tracking of non-cooperative unmanned aerial vehicles (UAVs) has
become a key requirement for safe operations in the low-altitude airspace~\cite{faa2020utm,
guvenc2018detection, azari2018key}. Unauthorized or malfunctioning UAVs may not report their
positions to a UAV traffic management system, yet they must be located and followed for safe
separation and, if needed, \rev{for} interdiction. A practical way to localize such a UAV is to listen
passively to the radio-frequency (RF) signals that it already emits, such as its command-and-control
and video links, with a network of time-synchronized ground sensors~\cite{guvenc2018detection}. Since the transmit time of a non-cooperative emitter is unknown, the time of arrival
(TOA) at a single sensor cannot be converted into a range. The time difference of arrival (TDOA)
between two sensors cancels the unknown transmit time, so the UAV position can be obtained by
multilateration from a set of TDOA measurements, with synchronization required only among the
ground sensors~\cite{ seyed2022handbook}.

The accuracy of TDOA-based UAV localization depends on the flight altitude and the signal bandwidth
and, most strongly, on whether each sensor has a line-of-sight (LOS) or a non-line-of-sight (NLOS)
link to the UAV. An NLOS link weakens the received signal and adds a positive excess delay, and the
resulting TDOA bias cannot be removed by averaging. Localization bounds, NLOS error
characterization, and UAV localization experiments have each been studied extensively, as reviewed
in Section~\ref{sec:lit}. However, an NLOS-aware Cram\'er--Rao lower bound (CRLB) analysis of aerial
TDOA localization that is evaluated against measurements, together with a model that reproduces the
measured errors, has not been reported.

In our earlier work~\cite{dickerson2025icc}, we derived a TDOA CRLB for LOS conditions, compared it
with measurements from five UAV flights on the Aerial Experimentation and Research Platform for
Advanced Wireless (AERPAW), and released the dataset~\cite{dickerson2025aerpaw}. That initial
evaluation reported the error together with the LOS states of the towers but did not model the NLOS
bias or the measured error distribution, and it did not process the position estimates. It also reported, without an explanation, a stable error of more than 30~m
while the UAV hovered at a waypoint where three of the four sensors were in LOS.

The specific contributions of this work, which is a rigorous extension
of~\cite{dickerson2025icc}, can be summarized as follows.
\begin{itemize}
\item We develop NLOS-aware CRLBs for 3D TDOA localization under two models of the NLOS biases. The
random-bias bound treats the biases as random with known statistics and includes them, together
with the excess attenuation of each NLOS link, in the TDOA covariance. The deterministic bound
treats the biases as unknown parameters of an augmented Fisher information matrix (FIM). A hybrid
bound with a Gaussian prior on the biases connects the two, as it equals the random-bias bound
when the prior matches the bias statistics and tends to the deterministic bound as the prior becomes
uninformative.
\item We show that the deterministic bound is finite only if the number of sensors exceeds the
number of NLOS links by at least four, or by three when the UAV altitude is known, so that knowing
the altitude is equivalent to one additional LOS sensor. For the AERPAW deployment, in which one
sensor is blocked throughout the flight, the position and the biases therefore cannot be separated
at any point in 3D and can be separated at only 45\% of the points when the altitude is known. We also quantify the
gains of additional LOS sensors and of relocating the blocked sensor.
\item \rev{We show that a localizer that ignores the NLOS biases, such as a weighted least-squares
localizer, is biased and has an error floor. To first order, the position offset that causes this
floor depends on the biases and the sensor geometry but not on the bandwidth, so a wider bandwidth
lowers the CRLB but not the floor.} The floor is consistent with the measured error, which remains
tens of meters above the CRLB wherever three sensors are in LOS and barely decreases between
\rev{2.5~MHz and 5~MHz}.
\item We develop a per-measurement error model that combines correlated shadowing, attitude and
antenna-pattern fluctuations, a heavy-tailed NLOS excess delay, mobility noise, and detection
outages. We estimate its seven unknown parameters jointly from the measured error distributions of
all flights by simulation-based minimum-distance estimation. The fitted model reproduces the median
errors of the four flights at \rev{2.5~MHz and 5~MHz} within 25\%, and its deviations in the error tails and
at 1.25~MHz point to the effects that it does not capture.
\item We characterize the measured error as a function of altitude, bandwidth, and the number of
LOS sensors, and evaluate outlier rejection\rev{,} and Kalman and particle filtering of the position
estimates. Outlier rejection based on the circular error probable (CEP) and a kinematic speed gate
lowers the pooled 90th-percentile error from \rev{850~m} to 99~m, whereas Kalman and particle filtering of
the retained fixes does not lower it further, because the residual error is dominated by bias.
\end{itemize}

The rest of this paper is organized as follows. Section~\ref{sec:lit} reviews the related
literature. In Section~\ref{sec:sys}, we introduce the system model and the LOS CRLB, and in
Section~\ref{sec:nlos}, we develop the NLOS-aware bounds and the bias-induced error floor. The
measurement model and the position estimator are presented in Section~\ref{sec:model}, followed by
the parameter estimation and post-processing procedures in Section~\ref{sec:est}.
Section~\ref{sec:exp} describes the AERPAW experiments and the dataset, and Section~\ref{sec:res}
presents the experimental and numerical results. Finally, Section~\ref{sec:conc} concludes the
paper, and the appendices contain the proofs.

\emph{Notation:} Vectors and matrices are denoted by bold lowercase and uppercase letters,
respectively. $(\cdot)^{\mathsf T}$ denotes the transpose, $\Tr(\cdot)$ the trace,
$\diag(\cdot)$ a diagonal matrix, $\E[\cdot]$ the expectation, $\lVert\cdot\rVert$ the Euclidean
norm, and $\mathcal N(\mu,\sigma^2)$ the Gaussian distribution. $\mathbf A\succeq\mathbf B$
($\mathbf A\succ\mathbf B$) means that $\mathbf A-\mathbf B$ is positive semidefinite (definite), and
$\mathbf 1$ and $\mathbf I$ denote the all-ones vector and the identity matrix.

\section{Literature Review}\label{sec:lit}
In this section, we review prior work on localization bounds and LOS TDOA methods, on NLOS error
bounds and mitigation, and on UAV localization experiments. Table~\ref{tab:lit} compares the scope of
the closest works with that of this paper.

\subsection{Localization Bounds and LOS TDOA Methods}
The CRLB is a standard tool to characterize the accuracy
of range-based localization~\cite{gezici2008survey, seyed2022handbook}. The impact of 3D antenna
radiation patterns on the CRLB of TDOA-based UAV localization is studied in~\cite{sinha2019impact},
and antenna-pattern effects on TOA-based 3D UAV localization, including a Gaussian NLOS bias on
links with random LOS states, are analyzed in~\cite{sinha2022impact}. The CRLB of 2D TOA localization with directive antennas is derived
in~\cite{khan2025cramer}, the TDOA localization performance of UAV-mounted sensors under flight
disturbances is analyzed in~\cite{8009702}, and the effect of the network topology and the cost
function on least-squares positioning is studied in~\cite{dizdarevic2006impact}. The CRLB also
serves as a design criterion and a benchmark. In~\cite{chen2025variable}, a variable-speed UAV path
is optimized for passive target localization by minimizing the CRLB, and in~\cite{kazzazi2025clear},
a closed-form estimator that combines TDOA and frequency difference of arrival and approaches the
CRLB with the minimum number of sensors is proposed. In~\cite{motie2024self}, a TDOA-based
self-localization method for UAVs without the global positioning system (GPS) is developed by least
squares with a convexity guarantee. However, apart from~\cite{sinha2022impact}, these works assume
LOS propagation, and none of them evaluates its bound against measurements.

\subsection{NLOS Error Bounds and Mitigation}
The CRLB of wireless geolocation in NLOS environments is
derived in~\cite{qi2006analysis}, where an NLOS measurement whose bias is unknown and has no prior is
shown to carry no position information; this result is generalized to wideband localization
in~\cite{shen2010fundamental}. A position error bound for ultra-wideband (UWB) localization in dense
cluttered environments is derived in~\cite{jourdan2008position}, and UWB geolocation techniques,
including the treatment of NLOS errors, are reviewed in~\cite{gezici2004uwb}. NLOS mitigation
algorithms based on residual weighting and on linear programming are proposed in~\cite{chen1999nlos}
and~\cite{venkatesh2006nlos}, respectively, and NLOS identification and mitigation for time-based
localization are surveyed in~\cite{nkrow2024nlos}. These studies consider terrestrial networks and
rely on analysis or simulation, and none of them evaluates an NLOS-aware bound with aerial TDOA
measurements.

\subsection{UAV Localization Experiments}
A UAV localization and tracking system based on passive RF
sensing and Kalman filtering is evaluated outdoors in~\cite{bhattacherjee2022experimental}, and a
scalable UWB TDOA system for the indoor navigation of multiple UAVs is presented in~\cite{8115937}.
However, these experiments neither compare the measured error with a localization bound nor model
the NLOS bias.

As Table~\ref{tab:lit} shows, each prior work covers only a subset of the five aspects. To the best
of our knowledge, this is the first work that combines an NLOS-aware CRLB analysis with a
measurement-validated error model for a deployed ground TDOA network that localizes UAVs in 3D, and
connects the two through a common set of fitted propagation parameters.

\begin{table}[!t]
\caption{Scope of closely related work versus this paper (\cmark: addressed; \pmark: partially
addressed; \xmark: not addressed). The columns indicate CRLB or FIM analysis, NLOS bias, a 3D or
UAV setting, real measurements, and error modeling.}
\label{tab:lit}
\centering
\footnotesize
\renewcommand{\arraystretch}{1.08}
\setlength{\tabcolsep}{4.5pt}
\begin{tabular}{lccccc}
\toprule
Work & CRLB & NLOS & 3D/UAV & Meas. & Model\\
\midrule
\cite{sinha2019impact, chen2025variable, kazzazi2025clear}
                                                   & \cmark & \xmark & \cmark & \xmark & \xmark\\
\cite{sinha2022impact}                             & \cmark & \cmark & \cmark & \xmark & \xmark\\
\cite{khan2025cramer}                              & \cmark & \xmark & \xmark & \xmark & \xmark\\
\cite{motie2024self}                               & \xmark & \xmark & \cmark & \xmark & \xmark\\
\cite{qi2006analysis, shen2010fundamental, jourdan2008position, gezici2004uwb}
                                                   & \cmark & \cmark & \xmark & \xmark & \xmark\\
\cite{chen1999nlos, venkatesh2006nlos, nkrow2024nlos}
                                                   & \xmark & \cmark & \xmark & \xmark & \xmark\\
\cite{bhattacherjee2022experimental, 8115937}      & \xmark & \xmark & \cmark & \cmark & \xmark\\
\cite{dickerson2025icc}                            & \cmark & \pmark & \cmark & \cmark & \xmark\\
\midrule
This work                                          & \cmark & \cmark & \cmark & \cmark & \cmark\\
\bottomrule
\end{tabular}
\end{table}

\section{System Model and LOS CRLB}\label{sec:sys}
In this section, we introduce the TOA and TDOA measurement model and the LOS CRLB, the baseline for
the NLOS analysis of Section~\ref{sec:nlos}. Table~\ref{tab:notation} summarizes the main symbols.

\begin{table}[!t]
\caption{Summary of the main notation.}
\label{tab:notation}
\centering
\footnotesize
\renewcommand{\arraystretch}{1.1}
\begin{tabular}{p{1.65cm}p{6.35cm}}
\toprule
Symbol & Definition\\
\midrule
$\x,\hat\x$ & True and estimated UAV position $[x,y,z]^{\mathsf T}$\\
\revt{$\x_{\mathrm h}$} & \revt{Horizontal UAV position $[x,y]^{\mathsf T}$}\\
$\x_i$, $N$ & Position of sensor $i$; number of sensors\\
$d_i$, $c\hat t_i$ & True and measured range to sensor $i$\\
$b_i$, $n_i$ & NLOS range bias ($b_i\ge0$) and noise of sensor $i$\\
$\sigma_i$, $\mathrm{SNR}_i$ & Ranging standard deviation and SNR of sensor $i$\\
$s_i$ & LOS state of sensor $i$ ($1$: LOS, $0$: NLOS)\\
$\beta$, $B$ & Effective bandwidth and noise bandwidth\\
$L_i$, $L_{\mathrm{ex}}$ & Excess-loss factor and NLOS excess loss (dB)\\
$r$, $i_\ell$ & Reference sensor; other sensor of the $\ell$th TDOA\\
$m$, $\bm\rho$ & Number of TDOAs ($m=N-1$); stacked TDOA vector\\
$\mathbf h(\x)$, $\mathbf J$ & Noise-free range differences and their Jacobian\\
$\mathbf Q$ & TDOA noise covariance\\
$\mathbf F$, $\mathbf C$ & FIM and CRLB covariance\\
$\mathcal K$, $\KN$ & Set of NLOS links and its cardinality\\
$\sigma_{b,i}$, $\mu_b$ & Bias standard deviation and median NLOS bias\\
$k$ & Measurement index (proxy for time)\\
\bottomrule
\end{tabular}
\end{table}

\subsection{TOA and TDOA Measurements}
\begin{figure}[!t]
\centering
\includegraphics[width=\columnwidth]{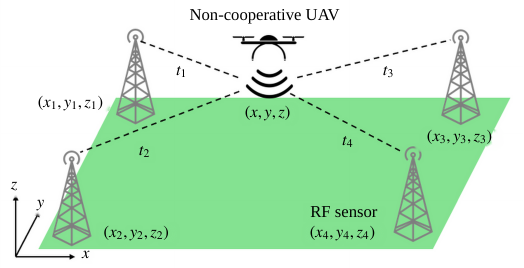}
\caption{Passive TDOA localization of a non-cooperative UAV at $(x,y,z)$ by time-synchronized RF
sensors at $(x_i,y_i,z_i)$, which measure the arrival times $t_i$ of its signal.}
\label{fig:sys}
\end{figure}

We consider $N$ passive, time-synchronized RF sensors at known fixed positions and a UAV at the
unknown position $\x=[x,y,z]^{\mathsf T}$, as illustrated in Fig.~\ref{fig:sys}. Sensor $i$ is
located at $\x_i=[x_i,y_i,z_i]^{\mathsf T}$, at the range $d_i=\lVert\x-\x_i\rVert$ from the UAV,
and forms a maximum-likelihood (ML) estimate $\hat t_i$ of the TOA of the UAV signal. By the
asymptotic normality of the ML estimator~\cite{seyed2022handbook}, the range equivalent of this
estimate can be written as
\begin{equation}
c\,\hat t_i = d_i + c\,t_0 + b_i + n_i, \qquad n_i\sim\mathcal N(0,\sigma_i^2),
\label{eq:toa}
\end{equation}
where $c$ is the speed of light, $t_0$ is the unknown transmit time, $b_i\ge0$ is the range bias
caused by an NLOS blockage, with $b_i=0$ on LOS links, and $n_i$ is the ranging noise. The NLOS bias
has been modeled as Gaussian~\cite{sinha2022impact}, exponential~\cite{gezici2004uwb}, or
uniform~\cite{venkatesh2006nlos}. The ranging standard deviation follows from the ML time-delay
estimation bound~\cite{gezici2008survey} as
\begin{equation}
\sigma_i=\frac{c}{2\sqrt{2}\,\pi\,\sqrt{\mathrm{SNR}_i}\,\beta},
\label{eq:sigma}
\end{equation}
where $\beta$ is the effective bandwidth and $\mathrm{SNR}_i$ is the received signal-to-noise ratio
(SNR) at sensor $i$.
With free-space path loss (FSPL), the SNR is given by
\begin{equation}
\mathrm{SNR}_i=\frac{P_{r,i}}{P_n},\ \
P_{r,i}=P_t G_t G_{r,i}\!\left(\frac{\lambda}{4\pi d_i}\right)^{\!2}\!L_i,\ \
P_n=k_B T B,
\label{eq:snr}
\end{equation}
where $P_t$ is the UAV transmit power, $G_t$ and $G_{r,i}$ are the transmit and receive antenna
gains, $\lambda$ is the wavelength, $k_B$ is Boltzmann's constant, $T$ is the temperature, $B$ is
the noise bandwidth, and $L_i\le1$ is an excess-loss factor, with $L_i=1$ on LOS links. We denote
the LOS state of sensor $i$ by $s_i\in\{0,1\}$, where $s_i=1$ indicates a link in LOS.

\rev{With} sensor $r$ as the reference, the TDOA measurements, expressed as range differences, are
$\rho_{r,i}=c(\hat t_r-\hat t_i)$ for $i\ne r$\rev{, which cancels the transmit time and can be expressed as}
\begin{equation}
\rho_{r,i}=h_{r,i}(\x)+(b_r-b_i)+(n_r-n_i),\qquad h_{r,i}(\x)=d_r-d_i.
\label{eq:tdoa}
\end{equation}
We denote by $i_\ell$ the non-reference sensor of the $\ell$th measurement. When all links are in
LOS, the vector $\bm\rho$ that stacks the $m=N-1$ measurements is Gaussian,
$\bm\rho\sim\mathcal N(\mathbf h(\x),\mathbf Q)$, where $\mathbf h(\x)$ stacks the range differences
$h_{r,i_\ell}(\x)$, and the covariance \rev{matrix can be expressed as}
\begin{equation}
\mathbf Q=\sigma_r^2\,\mathbf 1\mathbf 1^{\mathsf T}+\diag\big(\sigma_{i_1}^2,\dots,\sigma_{i_m}^2\big)\rev{,}
\label{eq:Q}
\end{equation}
\rev{which} has a shared-reference structure, since every TDOA contains the reference noise $n_r$, which
correlates all measurements through the common term $\sigma_r^2$.

\subsection{LOS Cram\'er--Rao Lower Bound}
The CRLB bounds the error covariance of any unbiased estimator $\hat\x$ of $\x$ as
$\E[(\hat\x-\x)(\hat\x-\x)^{\mathsf T}]\succeq\mathbf F^{-1}(\x)$, where the FIM is given
by~\cite{seyed2022handbook}
\begin{equation}
\mathbf F(\x)=\mathbf J^{\mathsf T}\mathbf Q^{-1}\mathbf J\in\Rt^{3\times3},
\qquad
\mathbf J=\frac{\partial\mathbf h(\x)}{\partial\x}\in\Rt^{m\times3},
\label{eq:fim}
\end{equation}
and, as is common for ranging-based localization, the information carried by the dependence of
$\mathbf Q$ on $\x$ is neglected. The $\ell$th row of $\mathbf J$ is the difference of two unit
vectors,
\begin{equation}
\frac{\partial h_{r,i_\ell}(\x)}{\partial\x}=\mathbf u_r-\mathbf u_{i_\ell},\qquad
\mathbf u_i=\frac{\x-\x_i}{\lVert\x-\x_i\rVert},
\label{eq:jac}
\end{equation}
where $\mathbf u_i$ points from sensor $i$ toward the UAV. From the CRLB covariance
$\mathbf C=\mathbf F^{-1}$, we obtain the 3D, vertical, and horizontal root-mean-square error (RMSE)
bounds
\begin{equation}
\begin{gathered}
\mathrm{RMSE}_{\mathrm{3D}}=\sqrt{\Tr(\mathbf C)},\qquad
\mathrm{RMSE}_{z}=\sqrt{C_{zz}},\\
\mathrm{RMSE}_{\mathrm h}=\sqrt{C_{xx}+C_{yy}},
\end{gathered}
\label{eq:rmse}
\end{equation}
where $C_{xx}$, $C_{yy}$, and $C_{zz}$ are the diagonal entries of $\mathbf C$. The horizontal bound
$\mathrm{RMSE}_{\mathrm h}$ is derived from the 3D CRLB, where altitude is estimated jointly with horizontal position. When the
altitude is known, as in our experiments \rev{described in Section~\ref{sec:exp}}, only
\rev{the horizontal position $\x_{\mathrm h}=[x,y]^{\mathsf T}$} is unknown, and the Jacobian reduces to its first two columns $\mathbf J_{xy}$. The altitude-known
horizontal bound is then given by
\begin{equation}
\mathrm{RMSE}_{\mathrm h}^{\mathrm{known}}=
\sqrt{\Tr\!\big[(\mathbf J_{xy}^{\mathsf T}\mathbf Q^{-1}\mathbf J_{xy})^{-1}\big]}
\le \mathrm{RMSE}_{\mathrm h},
\label{eq:chknown}
\end{equation}
where the inequality holds because knowing one parameter cannot reduce the information about the
others. In our setup, the sensors are mounted about 10~m above ground, and the UAV flies at
40--100~m at horizontal distances of several hundred meters. The elevation angles are therefore
small, the vertical geometry is poor, and $\mathrm{RMSE}_{\mathrm h}$ can be much larger than
$\mathrm{RMSE}_{\mathrm h}^{\mathrm{known}}$. Since the measured errors in this paper come from an
altitude-known solver, we compare them with $\mathrm{RMSE}_{\mathrm h}^{\mathrm{known}}$.

\section{NLOS-Aware CRLBs and the Bias-Induced Error Floor}\label{sec:nlos}
In this section, we extend the LOS bound to links that are blocked by terrain, foliage, or
buildings. An NLOS link degrades ranging in two ways. First, the blockage attenuates the received
power by the excess loss $L_{\mathrm{ex}}$ (in dB), which we model by setting
$L_i=10^{-L_{\mathrm{ex}}/10}$ in~\eqref{eq:snr} on each NLOS link; this lowers $\mathrm{SNR}_i$ and,
through~\eqref{eq:sigma}, increases $\sigma_i$. Second, the excess propagation delay adds the
positive bias $b_i$ in~\eqref{eq:toa} on the NLOS links, whose set we denote by
$\mathcal K=\{i:s_i=0\}$ and whose number we denote by $\KN=|\mathcal K|$; both depend on the UAV
position.

The resulting bound depends on what is known about the biases. In Model~A, the biases are random
with known statistics, which applies when NLOS statistics are available from ray tracing, a channel
model, or earlier measurements, but the individual biases are not calibrated. In Model~B, the biases
are deterministic unknowns, in which case the position and the biases may not be separable. A
hybrid bound connects the two models. All bounds assume that the LOS state of each link is known.

\subsection{Model A: Random-Bias CRLB}\label{sec:modelA}
We model the random part of each NLOS bias as a zero-mean Gaussian variable with variance
$\sigma_{b,i}^2$, independent across links; the bias mean is a deterministic offset that a
calibrated estimator can remove, and the effect of an uncompensated bias is analyzed in
Section~\ref{sec:floor}. Since the biases enter~\eqref{eq:tdoa} in the same way as the noise, they
add a covariance term with the shared-reference structure of~\eqref{eq:Q},
\begin{equation}
\begin{split}
\mathbf Q_{\mathrm{tot}}&=\mathbf Q_{\mathrm{noise}}+\mathbf Q_{\mathrm{bias}},\\
\mathbf Q_{\mathrm{bias}}&=\sigma_{b,r}^2\mathbf 1\mathbf 1^{\mathsf T}
+\diag\big(\sigma_{b,i_1}^2,\dots,\sigma_{b,i_m}^2\big),
\end{split}
\label{eq:Qtot}
\end{equation}
where $\mathbf Q_{\mathrm{noise}}$ is the noise covariance~\eqref{eq:Q} evaluated with the
attenuated NLOS SNRs, and $\sigma_{b,i}=0$ on LOS links. Replacing $\mathbf Q$ in~\eqref{eq:fim} by
$\mathbf Q_{\mathrm{tot}}$ yields the random-bias bound
\begin{equation}
\mathbf C_{\mathrm R}=\big(\mathbf J^{\mathsf T}\mathbf Q_{\mathrm{tot}}^{-1}\mathbf J\big)^{-1},
\label{eq:cB}
\end{equation}
which is finite for any non-degenerate geometry and captures both NLOS effects, the SNR loss
through $\mathbf Q_{\mathrm{noise}}$ and the bias uncertainty through $\mathbf Q_{\mathrm{bias}}$. For
non-Gaussian biases with the same variances, $\mathbf C_{\mathrm R}$ is the error covariance of the
best linear unbiased estimator and an upper bound on the exact CRLB, since the Gaussian distribution
has the smallest Fisher information for a given variance.

\subsection{Model B: Deterministic Unknown Biases}\label{sec:modelB}
When the bias statistics are unknown, the most conservative model treats each NLOS bias as a
deterministic unknown parameter that must be estimated together with the position. For TOA
localization, it is shown in~\cite{qi2006analysis, shen2010fundamental} that an NLOS measurement
whose bias is unknown and has no prior carries no position information. We extend this result to
the shared-reference TDOA model~\eqref{eq:tdoa} and express it as a condition on the number of
sensors. The unknown parameter vector becomes
\begin{equation*}
\bm\theta=\big[\,x,\ y,\ z,\ \{b_j\}_{j\in\mathcal K}\,\big]^{\mathsf T}\in\Rt^{3+\KN},
\end{equation*}
so that each NLOS link adds one unknown. By~\eqref{eq:tdoa}, the derivative of the $\ell$th TDOA
with respect to the bias $b_j$ is
\begin{equation}
\frac{\partial\rho_{r,i_\ell}}{\partial b_j}=\delta_{jr}-\delta_{j i_\ell},\qquad j\in\mathcal K,
\label{eq:dbias}
\end{equation}
where $\delta_{ab}$ denotes the Kronecker delta. A bias on the reference sensor thus shifts every
TDOA by $b_r$, whereas a bias on a non-reference sensor shifts only its own TDOA, by $-b_j$. With
the bias block $\mathbf J_b\in\Rt^{m\times\KN}$ built from~\eqref{eq:dbias} and the augmented
Jacobian $\mathbf J_{\mathrm a}=[\,\mathbf J\ \ \mathbf J_b\,]$, the augmented FIM and the position
block of its inverse are given by
\begin{equation}
\mathbf F_{\mathrm a}=\mathbf J_{\mathrm a}^{\mathsf T}\mathbf Q_{\mathrm{noise}}^{-1}\mathbf J_{\mathrm a},
\qquad
\mathbf C_{\mathrm D}=\big[\mathbf F_{\mathrm a}^{-1}\big]_{1:3,\,1:3},
\label{eq:cD}
\end{equation}
where $[\cdot]_{1:3,1:3}$ denotes the upper-left $3\times3$ block. The deterministic bound
$\mathbf C_{\mathrm D}$ accounts for the information that is spent on estimating the biases.

\begin{proposition}[Identifiability]\label{prop:ident}
The deterministic bound $\mathbf C_{\mathrm D}$ in~\eqref{eq:cD} is finite only if
\begin{equation}
m\ \ge\ 3+\KN\quad\Longleftrightarrow\quad N\ \ge\ 4+\KN .
\label{eq:ident}
\end{equation}
\rev{Conversely, the condition is sufficient whenever the geometry is non-degenerate, that is,
whenever any three rows $\mathbf u_r-\mathbf u_{i_\ell}$ of $\mathbf J$ that belong to LOS sensors are
linearly independent. This fails only for special arrangements, for example, when the UAV and all
LOS sensors lie in one plane.}
\end{proposition}
\begin{IEEEproof}
See Appendix~\ref{app:prop1}.
\end{IEEEproof}

Since $N-\KN$ is the number of LOS sensors, \eqref{eq:ident} requires at least four LOS sensors, as
many as 3D TDOA localization requires without NLOS links. Each unknown bias absorbs the TDOA that it
enters, which then no longer constrains the position. When
\eqref{eq:ident} holds, $\mathbf C_{\mathrm D}$ therefore equals the LOS bound of the LOS sensors
alone, and excluding the NLOS sensors loses no information but cannot restore identifiability
either, because it leaves $N-\KN$ unchanged. When the altitude is known, the same argument applies to
the two horizontal coordinates, and the condition becomes $N\ge3+\KN$, so that knowing the altitude
is equivalent to one additional LOS sensor. Section~\ref{sec:resB} evaluates both conditions for the
AERPAW deployment.

\subsection{Hybrid Bound}\label{sec:hybrid}
In practice, the biases are often partially known, for example, through a typical excess-delay
spread from a channel model or from earlier measurements. A bound that combines random parameters
with a prior and deterministic parameters without one is known as a hybrid
CRLB~\cite{rockah1987array}, and its purely random counterpart is the Bayesian
CRLB~\cite{vantrees2001detection}; bounds with a prior on the NLOS biases have been derived for TOA
localization in~\cite{qi2006analysis, jourdan2008position}. For the TDOA model, we place a
zero-mean Gaussian prior with variance $\sigma_{\mathrm{prior}}^2$ on each unknown bias. Since such
a prior adds the information $\sigma_{\mathrm{prior}}^{-2}$ to its own parameter
only~\cite{vantrees2001detection}, the hybrid FIM and the position bound are given by
\begin{equation}
\mathbf F_{\mathrm h}=\mathbf F_{\mathrm a}
+\diag\!\big(\mathbf 0_3,\ \sigma_{\mathrm{prior}}^{-2}\mathbf 1_{\KN}\big),
\qquad
\mathbf C_{\mathrm H}=\big[\mathbf F_{\mathrm h}^{-1}\big]_{1:3,\,1:3},
\label{eq:cH}
\end{equation}
where $\mathbf 0_3$ is the all-zeros vector of length three and $\mathbf 1_{\KN}$ the all-ones vector
of length $\KN$. Unlike $\mathbf F_{\mathrm a}$, the hybrid FIM $\mathbf F_{\mathrm h}$ is invertible
whenever $\mathbf J$ has rank three. Applying the Schur complement and the \rev{Woodbury identity}
to~\eqref{eq:cH} gives
\begin{equation}
\mathbf C_{\mathrm H}=\Big(\mathbf J^{\mathsf T}\big(\mathbf Q_{\mathrm{noise}}
+\sigma_{\mathrm{prior}}^2\,\mathbf J_b\mathbf J_b^{\mathsf T}\big)^{-1}\mathbf J\Big)^{-1},
\label{eq:cHclosed}
\end{equation}
in which the unknown biases act as additional noise with the covariance
$\sigma_{\mathrm{prior}}^2\mathbf J_b\mathbf J_b^{\mathsf T}$, which has the shared-reference
structure of $\mathbf Q_{\mathrm{bias}}$ in~\eqref{eq:Qtot}. Equation~\eqref{eq:cHclosed} connects the
three bounds. If all NLOS links have the same bias standard deviation $\sigma_b$ and
$\sigma_{\mathrm{prior}}=\sigma_b$, then $\mathbf C_{\mathrm H}$ equals the random-bias bound
$\mathbf C_{\mathrm R}$. As $\sigma_{\mathrm{prior}}\to0$, the biases become known, and
$\mathbf C_{\mathrm H}$ reduces to $(\mathbf J^{\mathsf T}\mathbf Q_{\mathrm{noise}}^{-1}\mathbf
J)^{-1}$, the LOS bound with the NLOS SNR loss. As $\sigma_{\mathrm{prior}}\to\infty$,
$\mathbf C_{\mathrm H}$ tends to the deterministic bound $\mathbf C_{\mathrm D}$, which is unbounded
whenever~\eqref{eq:ident} fails.

\subsection{Bias-Induced Error Floor}\label{sec:floor}
The bounds above limit the variance of unbiased estimators. \rev{However, a} practical localizer that
ignores the NLOS biases is \rev{a biased estimator}. To quantify this effect,
we write the stacked TDOA vector of~\eqref{eq:tdoa} as
\begin{equation}
\bm\rho=\mathbf h(\x)+\mathbf b_\rho+\mathbf n,\qquad
\mathbf n\sim\mathcal N(\mathbf 0,\mathbf Q_{\mathrm{noise}}),
\label{eq:rhomodel}
\end{equation}
where $\mathbf b_\rho$ stacks the TDOA biases $b_r-b_{i_\ell}$ and $\mathbf n$ stacks the noise
differences $n_r-n_{i_\ell}$. A localizer that does not know $\mathbf b_\rho$ computes the weighted
least-squares (WLS) estimate
\begin{equation}
\hat\x=\argmin_{\tilde\x}\
\big(\bm\rho-\mathbf h(\tilde\x)\big)^{\mathsf T}\mathbf W\big(\bm\rho-\mathbf h(\tilde\x)\big),
\label{eq:wls}
\end{equation}
where $\mathbf W\succ0$ is a weight matrix, and $\mathbf W=\mathbf Q_{\mathrm{noise}}^{-1}$ yields
the ML estimator without biases. For small errors, we linearize
$\mathbf h(\tilde\x)\approx\mathbf h(\x)+\mathbf J(\tilde\x-\x)$, with $\mathbf J$ as
in~\eqref{eq:fim}.

\begin{proposition}[Bias floor]\label{prop:floor}
Under the linearized model, the WLS estimate~\eqref{eq:wls} has the mean offset
\begin{equation}
\delta\x_{\mathrm{b}}=\E[\hat\x]-\x=\big(\mathbf J^{\mathsf T}\mathbf W\mathbf J\big)^{-1}
\mathbf J^{\mathsf T}\mathbf W\,\mathbf b_\rho,
\label{eq:pbias}
\end{equation}
and its mean-squared error (MSE) satisfies
\begin{equation}
\begin{split}
\mathrm{MSE}=\E\lVert\hat{\x}-\x\rVert^2
&=\lVert\delta\x_{\mathrm{b}}\rVert^2+\Tr\!\big(\mathrm{Cov}(\hat{\x})\big)\\
&\ge\lVert\delta\x_{\mathrm{b}}\rVert^2+\Tr(\mathbf C_{\mathrm n}),
\end{split}
\label{eq:msefloor}
\end{equation}
where $\mathbf C_{\mathrm n}=(\mathbf J^{\mathsf T}\mathbf Q_{\mathrm{noise}}^{-1}\mathbf J)^{-1}$,
with equality for $\mathbf W=\mathbf Q_{\mathrm{noise}}^{-1}$. Hence,
$\sqrt{\mathrm{MSE}}\ge\lVert\delta\x_{\mathrm{b}}\rVert$.
\end{proposition}
\begin{IEEEproof}
See Appendix~\ref{app:prop2}.
\end{IEEEproof}

Proposition~\ref{prop:floor} splits the error of a practical localizer into a variance term
$\Tr(\mathbf C_{\mathrm n})$, which is the CRLB of Section~\ref{sec:sys} evaluated with the
attenuated NLOS SNRs, and a bias term $\lVert\delta\x_{\mathrm b}\rVert^2$, which depends only on
the biases, on the geometry through $\mathbf J$, and on the relative link weights through
$\mathbf W$. The following corollary makes the bandwidth dependence of the two terms explicit.

\begin{corollary}[Bandwidth invariance of the floor]\label{cor:bw}
Let the effective bandwidth $\beta$ in~\eqref{eq:sigma} be proportional to the noise bandwidth $B$
in~\eqref{eq:snr}, and let the biases $\mathbf b_\rho$ be fixed. Then
$\mathbf Q_{\mathrm{noise}}(B)=(B_0/B)\,\mathbf Q_{\mathrm{noise}}(B_0)$ for a reference bandwidth
$B_0$, and the ML-weighted estimator ($\mathbf W=\mathbf Q_{\mathrm{noise}}^{-1}$) satisfies
\begin{equation}
\mathrm{MSE}(B)=\lVert\delta\x_{\mathrm b}\rVert^2
+\frac{B_0}{B}\,\Tr\big(\mathbf C_{\mathrm n}(B_0)\big)\ \xrightarrow{\;B\to\infty\;}\
\lVert\delta\x_{\mathrm b}\rVert^2,
\label{eq:msebw}
\end{equation}
where $\delta\x_{\mathrm b}$ does not depend on $B$.
\end{corollary}
\begin{IEEEproof}
See Appendix~\ref{app:prop2}.
\end{IEEEproof}

A wider bandwidth thus reduces the RMSE contribution of the variance term by a factor of $\sqrt2$
per doubling of $B$ but leaves the bias term unchanged, because a common scaling of $\mathbf W$
cancels in~\eqref{eq:pbias}. The $\sqrt2$ rate follows from the per-sample SNR in~\eqref{eq:snr};
if the receiver integrates the signal over a fixed time, $\sigma_i$ is proportional to $1/B$, the
RMSE contribution of the variance term halves per doubling of $B$, and the bias term is again
unchanged.
Corollary~\ref{cor:bw} assumes fixed biases; a wider bandwidth can also reduce the biases themselves
by resolving multipath, which is a separate effect. Section~\ref{sec:resC} compares these
predictions with the measurements.

\section{Measurement Model and Position Estimation}\label{sec:model}
In this section, we develop a per-measurement model that generates synthetic range measurements
along the measured trajectories, together with the position estimator that operates on them.
Section~\ref{sec:est} then estimates the model parameters from the measured errors.

\subsection{Per-Measurement Signal and Range Model}
Let $k$ denote the measurement index, which serves as a proxy for time, let $\mathcal T$ denote the
set of indices at which the UAV is in transit between waypoints, and let $\mathbb I\{A\}$ denote the
indicator function, which equals $1$ if the event $A$ occurs and $0$ otherwise. At index $k$, the
received SNR (in dB) at sensor $i$ is modeled as
\begin{equation}
\begin{split}
\mathrm{SNR}_i(k)=\ &\big[P_t+G_t+G_{r,i}-\mathrm{FSPL}(d_i(k))-N_{\mathrm F}-P_n\big]\\
&-L_{\mathrm{ex}}\,\big(1-s_i(k)\big)+X^{\mathrm{sh}}_i(k)+X^{\mathrm{att}}_i(k),
\end{split}
\label{eq:snrk}
\end{equation}
where the bracketed term is the free-space SNR of~\eqref{eq:snr} in dB, reduced by the receiver
noise figure $N_{\mathrm F}$; $P_t$ and $P_n$ are expressed in dBm, the antenna gains in dBi, and
$\mathrm{FSPL}(\cdot)$ in dB; and $s_i(k)$ is the LOS state at index $k$. The first random term
models correlated shadowing as a first-order autoregressive process,
\begin{equation}
X^{\mathrm{sh}}_i(k)=\psi\,X^{\mathrm{sh}}_i(k-1)+\sqrt{1-\psi^2}\,\sigma_{\mathrm{sh}}\,w_i(k),
\label{eq:shadow}
\end{equation}
where $w_i(k)\sim\mathcal N(0,1)$ is white Gaussian noise, and $\sigma_{\mathrm{sh}}$ equals
$\sigma_{\mathrm{sh}}^{\mathrm{LOS}}$ on LOS links and the larger
$\sigma_{\mathrm{sh}}^{\mathrm{NLOS}}$ on NLOS links. Following the Gudmundson
model~\cite{gudmundson1991correlation}, the correlation coefficient $\psi$ is close to one while
the UAV hovers, which freezes the shadowing, and smaller in transit, where the motion of the UAV
decorrelates it. The second random term $X^{\mathrm{att}}_i(k)\sim\mathcal N(0,\sigma_{\mathrm{att}}^2)$
models attitude and antenna-pattern fluctuations. Since roll, pitch, and yaw change the airframe
antenna gain toward each sensor~\cite{8690726}, $\sigma_{\mathrm{att}}$ is small in hover and larger
in transit.

The ranging standard deviation $\sigma_i(k)$ follows from~\eqref{eq:sigma} with the SNR
of~\eqref{eq:snrk}. Omitting the transmit time, which cancels in the TDOAs, the range measurement
at index $k$ is
\begin{equation}
\begin{split}
&\hat d_i(k)=d_i(k)+b_i(k)+b^{\mathrm{mp}}_i(k)+n_i(k),\\
&n_i(k)\sim\mathcal N\!\big(0,\ \sigma_i^2(k)+\sigma_{\mathrm{mob}}^2\,
\mathbb I\{k\in\mathcal T\}\big),
\end{split}
\label{eq:rangek}
\end{equation}
where $b_i(k)$ is the NLOS bias, which is log-normal with
$\log b_i\sim\mathcal N(\log\mu_b,\sigma_{\log}^2)$ on NLOS links, that is, with median $\mu_b$ and
log-standard deviation $\sigma_{\log}$, and zero on LOS links; $b^{\mathrm{mp}}_i(k)\sim\mathcal
N(0,\sigma_{\mathrm{mp}}^2)$ is a quasi-static multipath bias; and $\sigma_{\mathrm{mob}}$ captures
the additional ranging error in transit caused by the finite integration window and by residual
synchronization and Doppler errors. Both biases are frozen while the UAV hovers and are redrawn
independently at each index in transit. A sensor is detected only if $\mathrm{SNR}_i(k)$ exceeds
the threshold $\gamma_{\mathrm{th}}$, and no estimate is produced when fewer than three sensors are
detected. As a result, the model produces a nearly constant error offset in hover and a fluctuating
error in transit. Table~\ref{tab:modelparams} lists its parameters and the equations in which they
appear.

\begin{table}[!t]
\caption{Measurement-model parameters fitted to the raw and to the gated (outlier-removed)
measured errors\rev{, as discussed in Section~\ref{sec:resD},} and fixed parameters (SD: standard deviation; hover and
transit values are separated by a slash).}
\label{tab:modelparams}
\centering
\footnotesize
\renewcommand{\arraystretch}{1.08}
\setlength{\tabcolsep}{3.4pt}
\begin{tabular}{l>{\raggedright\arraybackslash}p{3.3cm}crr}
\toprule
Symbol & Description & Eq. & Raw & Gated\\
\midrule
\multicolumn{5}{l}{\emph{Estimated by error-distribution matching (Section~\ref{sec:fit})}}\\
$\mu_b$ (m) & NLOS bias median & \eqref{eq:rangek} & 54.1 & 61.5\\
$\sigma_{\log}$ & NLOS bias log-SD & \eqref{eq:rangek} & 1.01 & 1.41\\
$L_{\mathrm{ex}}$ (dB) & NLOS excess loss & \eqref{eq:snrk} & 13.1 & 14.5\\
$N_{\mathrm F}$ (dB) & Receiver noise figure & \eqref{eq:snrk} & 6.2 & 3.9\\
$\sigma_{\mathrm{sh}}^{\mathrm{NLOS}}$ (dB) & NLOS shadowing SD & \eqref{eq:shadow} & 6.9 & 8.1\\
$\sigma_{\mathrm{mp}}$ (m) & Multipath bias SD & \eqref{eq:rangek} & 5.8 & 6.8\\
$\tau_{\mathrm{res}}$ (m) & Residual rejection threshold & \eqref{eq:lm} & 31.6 & 18.5\\
\midrule
\multicolumn{5}{l}{\emph{Fixed}}\\
$\gamma_{\mathrm{th}}$ (dB) & Detection SNR threshold & \eqref{eq:snrk} & \multicolumn{2}{c}{$-5$}\\
$\psi$ & Shadowing correlation & \eqref{eq:shadow} & \multicolumn{2}{c}{0.97/0.6}\\
$\sigma_{\mathrm{att}}$ (dB) & Attitude fluctuation SD & \eqref{eq:snrk} & \multicolumn{2}{c}{1.5/3.5}\\
$\sigma_{\mathrm{sh}}^{\mathrm{LOS}}$ (dB) & LOS shadowing SD & \eqref{eq:shadow} & \multicolumn{2}{c}{4}\\
$\sigma_{\mathrm{mob}}$ (m) & Mobility ranging SD & \eqref{eq:rangek} & \multicolumn{2}{c}{6}\\
\bottomrule
\end{tabular}
\end{table}

\subsection{TDOA Position Estimation}
Like the geolocation software of the experimental system, the estimator operates in 2D with the
altitude known. Let $\mathcal D(k)$ denote the set of sensors detected at index $k$. With the
detected sensor of the highest SNR as the reference $r$, the Levenberg--Marquardt (LM) algorithm
solves
\begin{equation}
{\rev{\hat{\x}_{\mathrm h}(k)=\argmin_{\x_{\mathrm h}\in\Rt^2}}}\sum_{i\in\mathcal D(k)\setminus\{r\}}
\big[\rho_{r,i}(k)-h_{r,i}(\rev{\x_{\mathrm h}})\big]^2,
\label{eq:lm}
\end{equation}
where \rev{$\hat\x_{\mathrm h}(k)$ is the estimate of the horizontal position
$\x_{\mathrm h}=[x,y]^{\mathsf T}$ defined in Section~\ref{sec:sys}},
$\rho_{r,i}(k)=\hat d_r(k)-\hat d_i(k)$ is the TDOA formed from~\eqref{eq:rangek}, and
$h_{r,i}(\rev{\x_{\mathrm h}})$ is the range difference~\eqref{eq:tdoa} evaluated at
\rev{$\x=[\x_{\mathrm h}^{\mathsf T},z]^{\mathsf T}$} with the known altitude $z$. Equation~\eqref{eq:lm} is the
WLS estimator~\eqref{eq:wls} restricted to two dimensions with $\mathbf W=\mathbf I$. A solution
whose residual norm, divided by the number of TDOAs, exceeds the threshold $\tau_{\mathrm{res}}$ is
discarded as a no-estimate event; this gate models the consistency check of the geolocation
software.

\section{Model Parameter Estimation and Post-Processing}\label{sec:est}
In this section, we present two procedures that operate on position estimates. The first estimates
the unknown parameters of the measurement model from the measured errors, so that the model can
explain the error. The second processes the sequence of position estimates to remove gross outliers
and to smooth the remaining jitter.

\subsection{Measurement-Model Parameter Estimation by Error-Distribution Matching}\label{sec:fit}
Several parameters of~\eqref{eq:snrk}--\eqref{eq:lm}, such as the NLOS bias statistics, the excess
loss, and the receiver noise figure, are specific to the site and the hardware. They cannot be
measured directly from the released dataset, which contains only position estimates, yet they
determine how large and how heavy-tailed the localization error is. We therefore estimate the
seven parameters \rev{in Table~\ref{tab:modelparams}, namely}
\begin{equation}
\bm\phi=\{\mu_b,\ \sigma_{\log},\ L_{\mathrm{ex}},\ N_{\mathrm F},\
\sigma_{\mathrm{sh}}^{\mathrm{NLOS}},\ \sigma_{\mathrm{mp}},\ \tau_{\mathrm{res}}\}\rev{,}
\label{eq:phi}
\end{equation}
from the measured errors and fix the remaining parameters of Table~\ref{tab:modelparams} at typical
values to keep the number of unknowns small. The detection threshold $\gamma_{\mathrm{th}}$, for
example, enters the detection only through its sum with the estimated $N_{\mathrm F}$.

\begin{figure}[!t]
\centering
\includegraphics[width=\columnwidth]{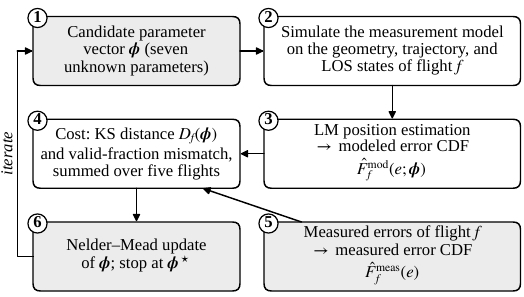}
\caption{Estimation of the model parameters by error-distribution matching. \rev{Step~1 selects a
candidate parameter vector $\bm\phi$. In steps~2 and~3,} the measurement model and the estimator are
simulated on the actual geometry, trajectory, and LOS states of each flight\rev{, and in step~4,} the
resulting error CDF is compared with the measured one \rev{of step~5} by the Kolmogorov--Smirnov (KS)
distance\rev{. In step~6,} the Nelder--Mead method updates $\bm\phi$ until the cost~\eqref{eq:ks} is
minimized.}
\label{fig:fitloop}
\end{figure}

The likelihood of the measured errors is intractable, because the error depends on $\bm\phi$
through the detection threshold, the LM solution, and the residual gate, but the model can be
simulated. We therefore use a simulation-based minimum-distance estimator in the spirit of
indirect inference~\cite{gourieroux1993indirect}, in which a statistic of the data is computed once
from the measurements and once from simulations of the model, and the parameters are adjusted until
the two agree, as illustrated in Fig.~\ref{fig:fitloop}\rev{, whose numbered steps are referred to
below}. We match the entire empirical error distribution
rather than a few moments, since the error is heavy-tailed and its moments are dominated by a few
extreme errors.

Let $f\in\{1,\dots,5\}$ index the flights, let $e_{f,k}$ denote the horizontal error of the $k$th
valid estimate of flight $f$, and let $n_f$ be the number of valid estimates. The empirical
cumulative distribution function (CDF) of the measured errors,
$\hat F^{\mathrm{meas}}_f(e)=n_f^{-1}\sum_{k=1}^{n_f}\mathbb I\{e_{f,k}\le e\}$, is the fraction of
estimates with an error of at most $e$. The modeled CDF
$\hat F^{\mathrm{mod}}_f(e;\bm\phi)$ is computed in the same way from $R_f$ Monte Carlo replicates
of flight $f$, each of which simulates~\eqref{eq:snrk}--\eqref{eq:lm} along the actual trajectory,
geometry, and per-sensor LOS states of \rev{flight $f$}. We estimate $\bm\phi$ \rev{in~\eqref{eq:phi} as}
\begin{equation}
\begin{split}
\bm\phi^\star=\argmin_{\bm\phi\in\Phi}\ \sum_{f=1}^{5}\Big[\,&D_f(\bm\phi)
+\eta\,\big|v_f^{\mathrm{meas}}-v_f^{\mathrm{mod}}(\bm\phi)\big|\,\Big],\\
D_f(\bm\phi)=\ &\sup_{e}\big|\hat F^{\mathrm{meas}}_f(e)-\hat F^{\mathrm{mod}}_f(e;\bm\phi)\big|,
\end{split}
\label{eq:ks}
\end{equation}
where $\Phi$ is the set of admissible (positive) parameter values, $D_f(\bm\phi)$ is the two-sample
Kolmogorov--Smirnov (KS) distance between the measured and modeled CDFs, $v_f$ is the fraction of measurement indices that
yield a valid estimate, and $\eta=0.6$ weights the two terms. \rev{Steps~2 and~3 of Fig.~\ref{fig:fitloop} compute
$\hat F^{\mathrm{mod}}_f$ and $v_f^{\mathrm{mod}}$, step~5 provides $\hat F^{\mathrm{meas}}_f$ and
$v_f^{\mathrm{meas}}$, and step~4 evaluates the cost in~\eqref{eq:ks}.}

The supremum in~\eqref{eq:ks} is taken over 220 logarithmically spaced error values between 1~m and
3.2~km. The KS distance depends only on the ranks of the errors, so a few extreme errors cannot
dominate the fit, but it is less sensitive in the tails than in the body of the distribution. The valid-fraction term constrains
$N_{\mathrm F}$ and $\tau_{\mathrm{res}}$, which mainly determine whether an estimate is produced.
Summing over all five flights forces a single parameter vector to reproduce every altitude and
bandwidth, which reduces the risk of overfitting a single flight. Since $D_f(\bm\phi)$ is piecewise
constant and non-convex, we minimize~\eqref{eq:ks} with the derivative-free Nelder--Mead simplex
method\rev{, which forms step~6 of Fig.~\ref{fig:fitloop},} starting from typical parameter values. Each
cost evaluation simulates $R_f=2$ replicates of each flight with fixed seeds, which makes the cost
deterministic in $\bm\phi$.

\subsection{Post-Processing: Outlier Rejection and Filtering}\label{sec:post}
\begin{figure}[!t]
\centering
\includegraphics[width=\columnwidth]{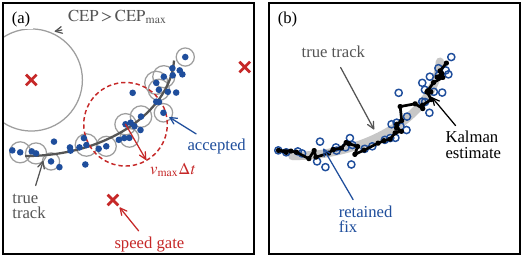}
\caption{Post-processing on a synthetic track. (a)~Fixes (dots) are shown with their CEPs (circles,
every third fix); fixes with a large CEP or an implausible jump from the last accepted fix are
rejected (crosses). \rev{The threshold examples show a fix with a CEP above
$\mathrm{CEP}_{\max}$ and a fix outside the speed-gate radius $v_{\max}\Delta t$ (dashed circle).} (b)~A random-walk Kalman filter (line with dots) smooths the
retained fixes (open circles). The gray line is the true track.}
\label{fig:concept}
\end{figure}

\rev{Like the estimator in~\eqref{eq:lm}, the geolocation software of the RF sensors produces one
position fix per measurement. With each fix, it also reports the circular error probable (CEP), which
is therefore available as part of the RF sensor measurements. The sequence of fixes contains two
types of error. A few fixes are gross outliers, hundreds of meters away from the true position,
which are caused by severe NLOS conditions or by a spurious multilateration solution. The other
fixes lie close to the true track but scatter around it from one fix to the next. Post-processing
therefore has two stages, which are illustrated in Fig.~\ref{fig:concept}. The first stage removes
the gross outliers, and the second stage smooths the retained fixes.}

\begin{figure*}[!t]
\centering
\begin{subfigure}[b]{0.30\textwidth}\centering
\includegraphics{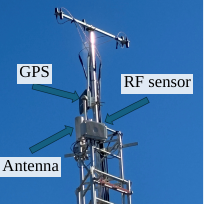}
\caption{RF sensor on tower LW3.}\label{fig:rfsensor}\end{subfigure}\hfill
\begin{subfigure}[b]{0.30\textwidth}\centering
\includegraphics{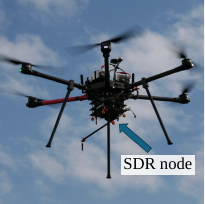}
\caption{UAV with the transmitting SDR.}\label{fig:uav}\end{subfigure}\hfill
\begin{subfigure}[b]{0.30\textwidth}\centering
\includegraphics[height=3.45cm,keepaspectratio]{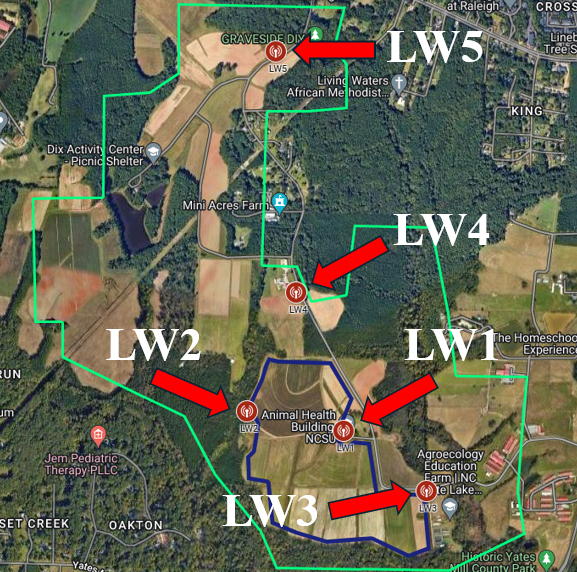}
\caption{Site and tower layout.}\label{fig:lwrfl}\end{subfigure}
\caption{AERPAW Lake Wheeler Field Road Labs site. (a)~Keysight N6841A RF sensor; the arrows point
to the sensor, its broadband antenna, and its GPS tracker. (b)~UAV carrying an AERPAW portable node
with a USRP B205mini SDR (arrow). (c)~Towers LW1--LW5; the RF sensors on LW2--LW5 provide the TDOA
measurements.}
\label{fig:aerpaw}
\end{figure*}

\rev{The first stage applies two tests, and a fix is rejected if it fails either of them. The first
test uses the CEP, which is the radius of a circle centered on the fix that contains the true
position with probability 0.5, so a large CEP indicates an unreliable fix. For a nearly circular
error distribution, $\mathrm{CEP}\approx0.59(\sigma_x+\sigma_y)$, where $\sigma_x$ and $\sigma_y$ are
the standard deviations of the east and north errors. The CEP gate rejects every fix whose CEP
exceeds the threshold $\mathrm{CEP}_{\max}$. The second test, a kinematic speed gate, checks whether
the UAV could have moved from the last accepted fix to the new fix. If the distance between the two
fixes, divided by the elapsed time $\Delta t$, exceeds the speed threshold $v_{\max}$, the jump is
physically implausible, and the new fix is rejected. In other words, a new fix is accepted only if it
lies within the radius $v_{\max}\Delta t$ around the last accepted fix. In Fig.~\ref{fig:concept}(a),
we show one example of each test. We set $v_{\max}$ well above the UAV speed, so that the scatter of
the fixes alone does not trigger the speed gate. If the TDOAs were available, random sample
consensus~\cite{fischler1981ransac} over subsets of sensors could detect the outliers in the
measurement domain instead.}

In the second stage, we filter the retained fixes with three filters: a constant-velocity (CV)
Kalman filter~\cite{barshalom2001estimation}, a random-walk (RW) Kalman filter, and an RW particle
filter with 500 particles and Student-$t$ measurement noise with three degrees of freedom, whose
heavy tails limit the influence of large errors~\cite{gustafsson2002particle}. The RW filters
propagate \rev{$\x_{\mathrm h,k}=\x_{\mathrm h,k-1}+\mathbf w_k$} with $\mathbf w_k\sim\mathcal N(\mathbf 0,q\,\Delta
t\,\mathbf I_2)$, where $q$ is the process-noise intensity and $\mathbf I_2$ is the $2\times2$
identity matrix, and use the CEP-adaptive measurement covariance
$\mathbf R_k=(\alpha\,\mathrm{CEP}_k)^2\mathbf I_2$, where the factor $\alpha$ is tuned, since the
reported CEP overstates the actual error. All filters use only the retained fixes.

\section{Experimental Setup and Dataset}\label{sec:exp}
In this section, we describe the AERPAW testbed, the RF sensors, the flights, and the dataset used
in Section~\ref{sec:res}.

The experiments were conducted at the AERPAW Lake Wheeler Field Road Labs site on July 15, 2024.
The sensors of Section~\ref{sec:sys} are four Keysight N6841A RF
sensors~\cite{keysight2021geolocation}\rev{, shown in Fig.~\ref{fig:rfsensor},} on towers LW2--LW5\rev{,
whose layout is shown in Fig.~\ref{fig:lwrfl}}. They detect and time-stamp RF signals between 20~MHz and 6~GHz, are
synchronized by GPS, and report to a common geolocation server, which estimates the 2D position of
the emitter with the altitude known and reports a CEP for each fix. During each flight, a USRP
B205mini software-defined radio (SDR) on the UAV\rev{, shown in Fig.~\ref{fig:uav},} transmitted a 3.32~GHz
channel-sounding waveform generated with GNU Radio, from which the sensors localized the UAV. The
transmit power was 30.67~dBm, the antenna gains were set to 0~dBi, and the temperature
in~\eqref{eq:snr} was $T=304.3$~K. This cooperative setup, with a known waveform and a known
altitude, is a favorable case for the passive localization of non-cooperative UAVs.

\begin{figure}[!t]
\centering
\includegraphics[width=\columnwidth]{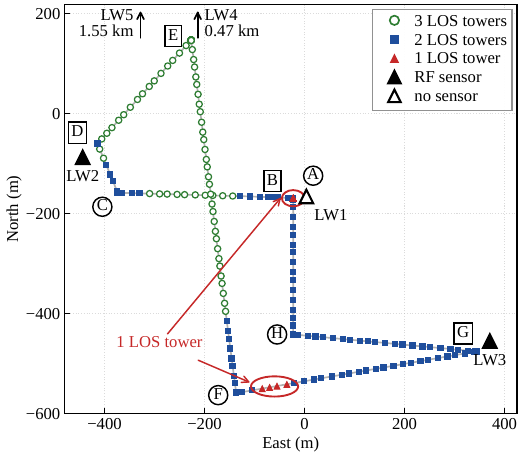}
\caption{Common UAV trajectory in local east--north coordinates, marked by the number of LOS towers
among LW2--LW5 during the 40~m flight. The UAV hovers for about 1~min at the waypoints in squares and
passes through those in circles. Filled triangles mark the RF sensors on LW2 and LW3, the open
triangle marks LW1, which has no sensor, and the arrows point to the sensors on LW4 and LW5,
\rev{0.47~km} and 1.55~km north of the origin. \rev{The ellipses mark the points with a single LOS
tower, at waypoint B and between waypoints F and G.}}
\label{fig:map}
\end{figure}

\rev{As listed in Table~\ref{tab:meas},} five flights, each flown once, cover altitudes of
\rev{40~m, 70~m,} and 100~m and bandwidths of \rev{1.25~MHz, 2.5~MHz,} and 5~MHz. In Fig.~\ref{fig:map}, we show the common closed path of the
flights through the waypoints A--H in local coordinates, together with the number of LOS towers
along the path at 40~m. The UAV flies at about 5~m/s between the waypoints and hovers for about
1~min at B, D, E, and G, and the system produces a position fix about every 4~s. Buildings and
trees were modeled in the Java OpenStreetMap editor and imported into MATLAB for ray tracing, which
yields the LOS state of each link along the trajectory; satellite views of these LOS states at the
three altitudes are given in~\cite{dickerson2025icc}. Tower LW5, about 1.5~km north of the field,
is in NLOS at every point of every flight, and tower LW1, which is inside the flight area but did not
host an RF sensor, is used in the placement study of Section~\ref{sec:resB}.

A measurement without a returned position counts as a no-estimate event. The released
dataset~\cite{dickerson2025aerpaw} contains the position estimates, the ground truth, the per-tower
LOS states, and the CEP of each fix, but not the TDOAs. Our processing follows the released MATLAB
script and reproduces the error statistics of~\cite{dickerson2025icc} exactly\rev{, as shown in
Table~\ref{tab:meas}}.

\begin{table}[!t]
\caption{Measured horizontal localization error per flight (valid fixes; P90: 90th
percentile).}
\label{tab:meas}
\centering
\footnotesize
\setlength{\tabcolsep}{5pt}
\begin{tabular}{lrrrrr}
\toprule
Flight & Valid/ & Valid & Mean & Median & P90\\
       & total  & (\%)  & (m)  & (m)    & (m)\\
\midrule
40 m, 1.25 MHz & 186/255 & 72.9 & 312.0 & 117.1 & 891.5\\
40 m, 2.5 MHz  & 168/242 & 69.4 & 280.3 &  59.0 & 980.0\\
40 m, 5 MHz    & 185/245 & 75.5 & 358.8 &  51.6 & 1132.0\\
70 m, 5 MHz    & 205/245 & 83.7 & 256.4 &  42.9 & 666.0\\
100 m, 5 MHz   & 240/261 & 92.0 & 163.6 &  33.2 & 312.0\\
\bottomrule
\end{tabular}
\end{table}

\begin{figure*}[!t]
\centering
\includegraphics[width=\textwidth]{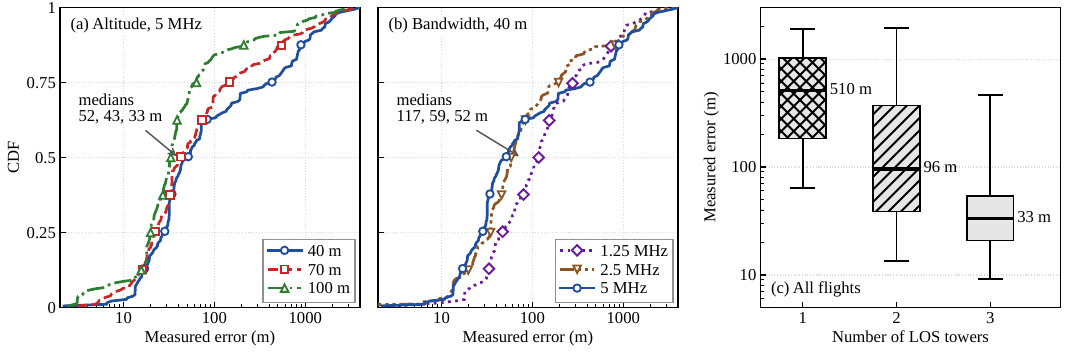}
\caption{Measured horizontal localization error. (a)~CDFs for the three altitudes at 5~MHz and
(b)~for the three bandwidths at 40~m, with the medians annotated. (c)~Error of all flights grouped by
the number of LOS towers among LW2--LW5 (40, 476, and 468 fixes with one, two, and three LOS
towers), with the median next to each box (boxes: interquartile range; whiskers: 5th and 95th
percentiles).}
\label{fig:realcdf}
\end{figure*}

\section{Experimental and Numerical Results}\label{sec:res}
In this section, we first characterize the measured error and then evaluate the NLOS-aware bounds,
the measurement model, and the post-processing on the AERPAW data. Unless otherwise noted, the
bounds are evaluated at 5~MHz along the measured trajectories with the actual sensor positions and
per-link LOS states. The NLOS parameters are the fitted values of Table~\ref{tab:modelparams}, and
the random-bias bound~\eqref{eq:cB} uses $\sigma_{b,i}=54$~m, the fitted median bias $\mu_b$; since the fitted
log-normal bias has a larger spread, this choice can only lower the bound. As
in~\cite{dickerson2025icc}, we set $\beta=B$ in~\eqref{eq:sigma} and use the per-sample SNR
of~\eqref{eq:snr}, and all bounds include the fitted noise figure $N_{\mathrm F}$. The
root-mean-square bandwidth $B/\sqrt{12}$ of a flat spectrum would raise the variance-limited bounds
by a factor of 3.5, and integrating over $n$ samples would lower them by $\sqrt n$, but neither
choice affects the bias-dominated bounds.
Horizontal bounds are given by~\eqref{eq:chknown} unless they are labeled as 3D, in which case they
follow from~\eqref{eq:rmse}.

\begin{figure*}[!t]
\centering
\includegraphics[width=\textwidth]{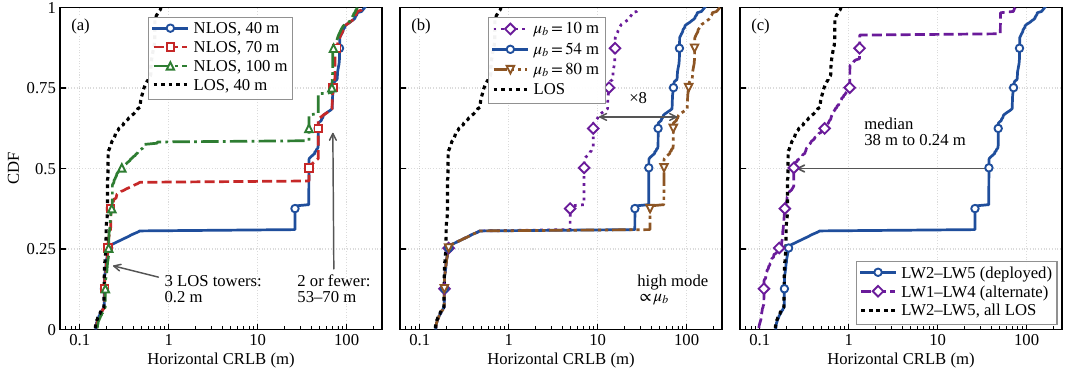}
\caption{CDFs of the altitude-known horizontal CRLB along the trajectory at 5~MHz. (a)~Random-bias
NLOS bound under the actual LOS states for the three altitudes and LOS bound at 40~m; the arrows mark
the two modes. (b)~NLOS bound at 40~m for three bias levels, with $\sigma_{b,i}=\mu_b$, and LOS
bound; the arrow marks the factor between $\mu_b=10$~m and \rev{$\mu_b=80$~m} in the high mode.
(c)~Deployed network under the actual LOS states, \rev{an alternate} network in which LW5 is replaced by
LW1, assumed in LOS, and the
deployed network with all links in LOS; the arrow marks the change of the median.}
\label{fig:bounds}
\end{figure*}

\subsection{Effect of Altitude, Bandwidth, and LOS Conditions}\label{sec:resA}
In Fig.~\ref{fig:realcdf}, we show the CDFs of the measured horizontal error for the three altitudes
at 5~MHz and the three bandwidths at 40~m, together with the error of all flights grouped by the
number of LOS towers. \rev{As shown in Fig.~\ref{fig:realcdf}(a),} the median error decreases with
altitude, from 52~m at 40~m to 43~m at 70~m and 33~m at 100~m. \rev{As shown in
Fig.~\ref{fig:realcdf}(b), it also decreases} with bandwidth, from 117~m at 1.25~MHz to 59~m at
2.5~MHz and 52~m at 5~MHz. \rev{As seen in Table~\ref{tab:meas}, however,} the mean and the 90th
percentile do not decrease with bandwidth, and altitude also raises the fraction of valid estimates,
from 75.5\% at 40~m to 92.0\% at 100~m. \rev{In Fig.~\ref{fig:realcdf}(c), the} number of LOS towers
has an even larger effect, as the median error rises from 33~m with three LOS towers to 96~m with two
and 510~m with one\rev{. O}ver the valid fixes, LW2, LW3, and LW4 are in LOS 92\%, 97\%, and
54\% of the time, respectively, and LW5 never. The altitude dependence is a consequence of this
effect, because a higher altitude clears more of the trees and buildings around the field and thus
adds LOS towers, as quantified in Section~\ref{sec:resB}.

\subsection{NLOS CRLBs, Identifiability, and Sensor Placement}\label{sec:resB}
\rev{Fig.~\ref{fig:bounds} shows the CDFs of the altitude-known horizontal CRLB along the
trajectory at 5~MHz, and we discuss its three panels in turn.} \rev{Fig.~\ref{fig:bounds}(a) compares the random-bias NLOS bound under the actual LOS states for the
three altitudes, together with the LOS bound at 40~m. The LOS bound ranges from 0.15 to 0.8~m, whereas the NLOS bound is bimodal. With three LOS towers, the NLOS bound remains below 1~m (median 0.2~m), close to the LOS
bound. With fewer than three towers in LOS, its median rises to 53--70~m, over two orders of magnitude above the LOS value. This is because three LOS towers determine the
horizontal position without the blocked link, whereas with two LOS towers, the position also depends
on a TDOA with an uncertain bias. Increasing altitude raises the fraction of points in the low mode, from 31\% at 40~m to 46\% at 70~m and 58\% at 100~m, consistent with the measured altitude
dependence in Section~\ref{sec:resA}. Table~\ref{tab:crlb} lists the
corresponding medians for the three altitudes, separately at the points with three and with at most
two LOS towers.}

\rev{Fig.~\ref{fig:bounds}(b), shows the 40~m NLOS bound for three bias levels, with
$\sigma_{b,i}=\mu_b$. The low mode barely changes, while the high mode grows nearly proportionally to $\mu_b$, increasing eightfold from $\mu_b=10$~m to 80~m. The
high mode is therefore governed by the bias uncertainty rather than by the thermal noise.}

\rev{Fig.~\ref{fig:bounds}(c) compares the deployed network with one replacing LW5 by LW1 inside the flight area Fig.~\ref{fig:map}. Assuming
LW1 is always in LOS, the 40~m median bound falls from 38~m to
0.24~m and the 90th percentile from 83~m to 1.3~m, approaching the all-LOS bound. Replacing the always-blocked LW5 preserves three LOS towers even when one of LW2--LW4 is blocked. Since the field-wide LOS state
of LW1 was not ray traced, this gain is an optimistic estimate.}

\begin{table}[!t]
\caption{Medians of the horizontal and vertical CRLBs along the trajectory at 5~MHz, with the 90th
percentiles of the larger bounds in brackets. The LOS bound is taken over all points, and the
altitude-known NLOS bound separately over the points with three and with at most two LOS towers.}
\label{tab:crlb}
\centering
\footnotesize
\setlength{\tabcolsep}{3.3pt}
\begin{tabular}{lrrrrrr}
\toprule
 & Points & \multicolumn{3}{c}{Altitude known (m)} & \multicolumn{2}{c}{3D localizer, NLOS}\\
\cmidrule(lr){3-5}\cmidrule(lr){6-7}
Alt. & with & LOS & NLOS: & NLOS: & Horiz. & Vert.\\
 & 3 LOS & & 3 LOS & $\le$2 LOS & (m) & (km)\\
\midrule
40 m  & 31\% & 0.21 & 0.19 & 53 [91]  & 146 [724] & 2.6\\
70 m  & 46\% & 0.22 & 0.20 & 70 [99]  & 143 [686] & 1.1\\
100 m & 58\% & 0.21 & 0.21 & 70 [110] & 123 [615] & 0.83\\
\bottomrule
\end{tabular}
\end{table}

The 3D columns of Table~\ref{tab:crlb} quantify the cost of not knowing the altitude, which is the
case for a non-cooperative UAV. Since the sensors view the UAV at small elevation angles, the median
vertical bound is 0.8--2.6~km, and the altitude uncertainty raises the median horizontal bound to
123--146~m.

\begin{remark}[Identifiability of the AERPAW deployment]\label{rem:aerpaw}
Since LW5 is blocked at every point, $\KN\ge1$ everywhere, and the deployed network with $N=4$
violates~\eqref{eq:ident} at every point. A 3D localizer therefore cannot separate the position
from the NLOS biases anywhere, and $\mathbf C_{\mathrm D}$ is unbounded. With the altitude known, the
condition $N\ge3+\KN$ holds only where LW5 is the only blocked tower, at 45\% of the points of the
three 5~MHz flights. Identifiability requires additional LOS sensors, added or relocated, or prior
information on the biases, as in Model~A.
\end{remark}

\begin{figure}[!t]
\centering
\includegraphics[width=\columnwidth]{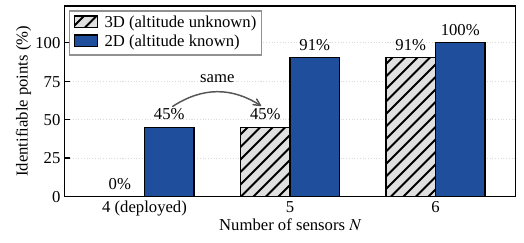}
\caption{Fraction of trajectory points of the three 5 MHz flights at which the deterministic-bias CRLB is finite, versus the number of sensors $N$. The additional sensors are in LOS; their positions do not affect identifiability as long as the geometry is non-degenerate. The arrow marks that knowing the altitude has the same effect as one additional sensor.}
\label{fig:ident}
\end{figure}

In Fig.~\ref{fig:ident}, we show the fraction of trajectory points at which the deterministic-bias
CRLB is finite as a function of the number of sensors, for a 3D localizer and with the altitude
known. One and two additional LOS sensors raise the 3D fraction from zero to \rev{45\% and 91\%,
respectively,} the
fractions reached with the altitude known and one sensor fewer. This is because each additional LOS
sensor raises $N-\KN$ by one, so that, as Proposition~\ref{prop:ident} predicts, knowing the
altitude has the same effect as one additional LOS sensor.

\begin{figure}[!t]
\centering
\includegraphics[width=\columnwidth]{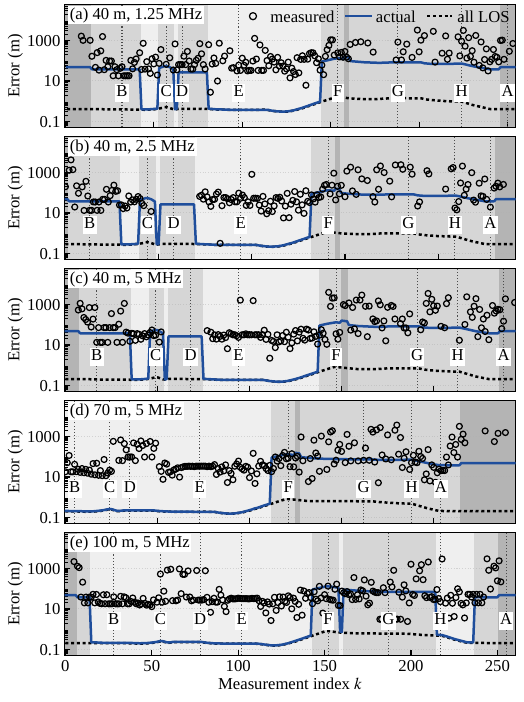}
\caption{Measured error of each fix (circles) versus the measurement index $k$ for the five flights,
with the altitude-known horizontal CRLB at the bandwidth of each flight under the actual LOS states
(solid) and with all links in LOS (dotted). The shading gives the number of LOS towers among
LW2--LW5 (light: three, medium: two, dark: one), and the vertical dotted lines and letters mark the
waypoints.}
\label{fig:errk}
\end{figure}

\subsection{Per-Measurement Error and the Bias Floor}\label{sec:resC}
Fig.~\ref{fig:errk} shows the measured error per fix versus measurement
index $k$ for all five flights, alongside the altitude-known CRLB under actual and all-LOS conditions. The measured error follows the bound's two modes but never
reaches the lower one. Errors are largest with at most two
LOS towers, particularly near waypoints F, G, and H, where
the bound is tens of meters. With three LOS towers, the
bound is below 1~m, but errors remain at tens of meters,
except for clusters of several-hundred-meter errors between
C and D at 70~m and 100~m.

This gap is consistent with the bias floor of Proposition~\ref{prop:floor}. If the geolocation
software follows~\eqref{eq:lm}, it weights all TDOAs equally($\mathbf W=\mathbf I$), so LW5's excess delay shifts
the estimates, whereas the random-bias bound assumes weights that effectively
discard LW5. At points with three LOS towers,
\eqref{eq:pbias} with $\mathbf W=\mathbf I$ and
$\mu_b=54$~m on LW5 predicts a median offset
$\lVert\delta\x_{\mathrm b}\rVert$ of 19--27~m, comparable
to the measured 28--40~m at \rev{2.5~MHz and 5~MHz}.
Since~\eqref{eq:pbias} is linear in $\mathbf b_\rho$,
halving the bias to 27~m reduces the predicted offset
to 10--14~m.

The measured median's bandwidth dependence agrees with
Corollary~\ref{cor:bw}, while gross errors dominate the
mean and 90th percentile. From \rev{2.5~MHz to 5~MHz},
the 40~m median falls only from \rev{59~m} to 52~m,
versus about 42~m predicted for variance-limited error
using the per-sample SNR of~\eqref{eq:snr}, or 30~m
with fixed integration time. The larger reduction from
\rev{117~m} to 59~m between \rev{1.25~MHz and 2.5~MHz}
indicates unmodeled narrowband degradation
(Section~\ref{sec:resD}). The bias floor can be reduced
by excluding or de-weighting NLOS links or improving
sensor geometry.

The waypoint structure also clarifies the anomaly reported in~\cite{dickerson2025icc}. While the UAV
hovers at waypoint E, where only LW5 is blocked, erros remain at 32--34~m across all three 5~MHz
flights, with an interquartile range below 4~m. At this location, the LW5 bias causes an offset of
only about 6~m. By~\eqref{eq:pbias}, a static bias of 40--55~m (130--180~ns) on any LOS link would produce the observed error, possibly from an unresolved reflection off a nearby structure~\cite{coutts2000passive} or sensor timing offset. The model of Section~\ref{sec:model} reproduces the freezing of such a bias
during hover, but with the fitted $\sigma_{\mathrm{mp}}$, a static bias of this size on a LOS link is
improbable, which points to an unmodeled site-specific effect.

\begin{figure*}[!t]
\centering
\includegraphics[width=\textwidth]{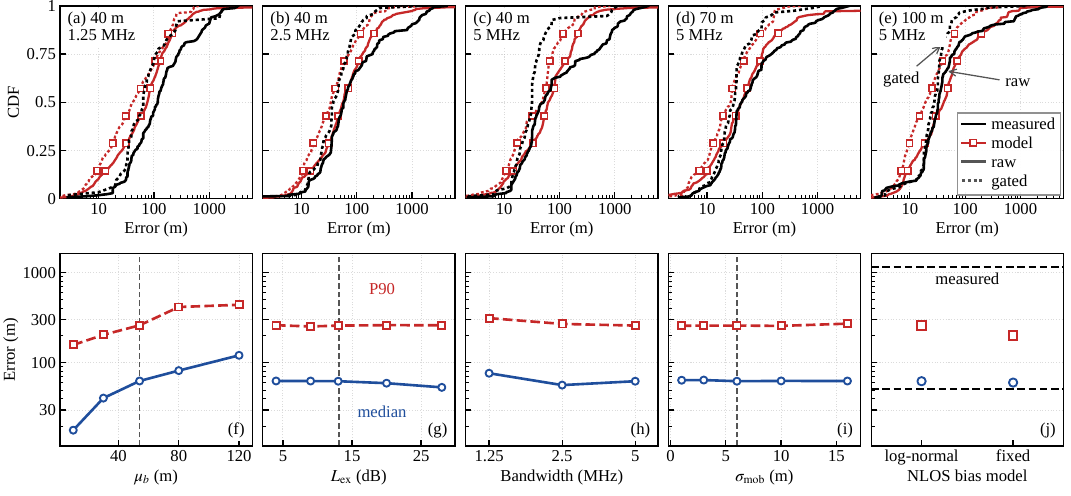}
\caption{Measurement model \rev{performance}. (a)--(e)~Measured (no markers) and modeled (squares) error CDFs, raw
(solid) and gated, that is, after outlier rejection (dotted), with the raw and the refitted parameters
of Table~\ref{tab:modelparams}; the CEP gate applies only to the measured fixes. (f)--(i)~Modeled
median (circles) and 90th-percentile (squares) errors of the 40~m, 5~MHz flight versus four
parameters, with the fitted or nominal values dashed; (h) uses the three 40~m flights. (j)~Same
errors for a log-normal and a fixed NLOS bias, with the measured values dashed.}
\label{fig:fit}
\end{figure*}

\begin{figure*}[!t]
\centering
\includegraphics[width=\textwidth]{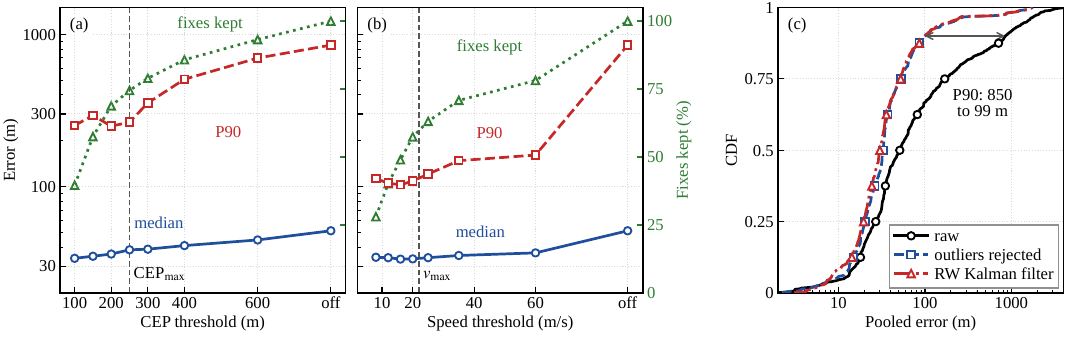}
\caption{Post-processing of the measured fixes. (a)~CEP-gate and (b)~speed-gate sweeps of the
pooled median (circles) and P90 (squares) errors and of the fraction of fixes kept (triangles, right
axis), with the selected thresholds dashed. (c)~Pooled error CDFs of the raw fixes, after outlier
rejection, and after RW Kalman filtering; the arrow marks the P90 reduction.}
\label{fig:post}
\end{figure*}

\begin{table}[!t]
\caption{Measured versus modeled error per flight for the raw fit.
\rev{Modeled errors use \eqref{eq:snrk}--\eqref{eq:lm}
with parameters estimated via \eqref{eq:ks}
(six Monte Carlo replicates).}}
\label{tab:fit}
\centering
\footnotesize
\setlength{\tabcolsep}{4.2pt}
\begin{tabular}{lrrrrr}
\toprule
 & KS & \multicolumn{2}{c}{Median (m)} & \multicolumn{2}{c}{P90 (m)}\\
\cmidrule(lr){3-4}\cmidrule(lr){5-6}
Flight & dist. & Meas. & Model & Meas. & Model\\
\midrule
40 m, 1.25 MHz & 0.19 & 117 & 77 & 891  & 311\\
40 m, 2.5 MHz  & 0.10 &  59 & 57 & 980  & 269\\
40 m, 5 MHz    & 0.21 &  52 & 63 & 1132 & 258\\
70 m, 5 MHz    & 0.12 &  43 & 42 & 666  & 304\\
100 m, 5 MHz   & 0.15 &  33 & 41 & 312  & 343\\
\bottomrule
\end{tabular}
\end{table}

\subsection{Measurement-Model Fit and Parameter Sensitivity}\label{sec:resD}
Figs.~\ref{fig:fit}(a)--(e) compare the measured error CDFs of five flights with \rev{those generated by the measurement model~\eqref{eq:snrk}--\eqref{eq:lm},}
before and after outlier rejection. Table~\ref{tab:fit} reports the KS distances, medians, and 90th percentiles of the raw errors. Table~\ref{tab:modelparams} lists the parameters
fitted via~\eqref{eq:ks}, which are shared across all flights; only altitude, bandwidth, trajectory, and LOS states vary. For the four flights at
\rev{2.5~MHz and 5~MHz}, modeled medians are within 25\% of measurements, with KS distances of 0.10 to 0.21. However, the model underestimates the 90th percentile at \rev{40~m and 70~m}
by factors of 2.2 to 4.4\rev{. A}t 1.25~MHz, it underestimates the error throughout, with a median of 77~m versus the
measured 117~m. This is because the model lacks gross-error mechanisms, such as spurious
multilateration solutions, which would produce the long measured tails, while~\eqref{eq:sigma}
captures neither ML delay ambiguity near the detection threshold nor unresolved
multipath at coarse time resolution.
Thus, the model reproduces the typical errors but the heaviest tails or narrowband degradation.

Outlier rejection removes most errors above 100~m and
lowers the median in both measured and modeled data,
while retaining about half the fixes. Because the model
does not produce CEP, only the speed gate applies to
modeled fixes, yielding different rejection behavior.
Refitting $\bm\phi$ after outlier rejection changes the
median NLOS bias by 14\% (\rev{54~m} to 62~m), but
$\sigma_{\log}$, $N_{\mathrm F}$, and $\tau_{\mathrm{res}}$
by about 40\% \rev{(Table~\ref{tab:modelparams})}.
The latter parameters trade off with each other and
the missing CEP gate, while the median bias remains stable.

Figs.~\ref{fig:fit}(f)--(i) show the modeled median and 90th-percentile errors versus NLOS bias level $\mu_b$, excess loss $L_{\mathrm{ex}}$,
bandwidth, and mobility noise $\sigma_{\mathrm{mob}}$ for the 40~m, 5~MHz flight, with the other parameters fixed at fitted
or nominal values. Raising $\mu_b$ from \rev{10~m} to 120~m increases the modeled median from
\rev{18~m} to 121~m and the 90th percentile from \rev{158~m} to 437~m. In contrast, $L_{\mathrm{ex}}$ and $\sigma_{\mathrm{mob}}$
change the median by less than 15\%, while the modeled median across the three 40~m flights varies between
\rev{57~m} and 77~m with the bandwidth. As Proposition~\ref{prop:floor} shows, NLOS bias sets the offset that dominates the error, whereas a larger excess loss mainly pushes NLOS links below the
detection threshold, which removes them from the solution together with their biases and slightly
lowers the median. Fig.~\ref{fig:fit}(j) shows that the spread of the bias matters as
well, since a fixed bias of $\mu_b$ lowers the modeled 90th percentile from \rev{258~m} to 200~m,
further below the measured 1132~m. The model must therefore capture the NLOS bias statistics; the other
parameters matter less.

\subsection{Outlier Rejection and Filtering}\label{sec:resE}
\rev{Figs.~\ref{fig:post}(a) and (b)} show the pooled median and
90th-percentile errors and fraction of fixes kept as a function of the CEP and
speed thresholds\rev{, respectively}. CEP indicates fix quality, with a rank correlation of 0.40 with error. Fixes with CEP below 200~m have a median error of 36~m versus
168~m otherwise. \rev{In Fig.~\ref{fig:post}(b), the} speed gate alone lowers the
90th-percentile error from \rev{850~m} to 109~m at $v_{\max}=20$~m/s retaining 57\% of the fixes.
Gross outliers produce jumps of hundreds of meters, implying unrealistic speeds far above the UAV's 5~m/s. With
$\mathrm{CEP}_{\max}=250$~m and $v_{\max}=22$~m/s, chosen by inspecting these sweeps, outlier
rejection retains 52\% of the fixes. \rev{ Fig.~\ref{fig:post}(c), compares the pooled error CDFs before and after outlier rejection, and after RW Kalman filtering. Outlier rejection} lowers the
pooled median error from \rev{51~m} to 33~m and the 90th-percentile error from \rev{850~m} to 99~m.
This in-sample gain does not hinge on the exact thresholds, as the 90th percentile stays at
103--121~m for speed thresholds of \rev{8~m/s} to 25~m/s.

Filtering offers little additional benefit. An RW Kalman filter with weak smoothing
($q=100$~m$^2$/s, $\alpha=0.15$) reduces the median to 30~m but raises the 90th percentile from \rev{99~m} to 102~m. Stronger smoothing ($q=4$~m$^2$/s) lags the stop-and-go motion, increasing the
median to 53~m and the 90th percentile to 190~m. A CV Kalman filter reaches a similar median but raises the 90th
percentile to 110--116~m due to velocity drift across gaps in the 4~s track. The particle filter occasionally loses track, raising the 90th percentile to 771~m. After outlier rejection, the
remaining error is dominated by the bias floor of Section~\ref{sec:resC}, which position-domain
filtering cannot remove. Thus, outlier rejection is the most effective
post-processing step.

\section{Conclusion}\label{sec:conc}
In this paper, we developed NLOS-aware CRLBs and an empirically calibrated measurement model for
TDOA-based UAV localization, evaluated using five AERPAW flights. With known altitude, the random-bias bound is below 1~m with three LOS towers but over two
orders of magnitude larger otherwise, consistent with the measured altitude dependence. The
identifiability condition shows that the deployed network cannot separate the position from the
NLOS biases at any point in 3D and that knowing the altitude is equivalent to one additional LOS
sensor. The measured error remains tens of meters above the bound and barely decreases between
\rev{2.5~MHz and 5~MHz}, as the bias-induced error floor predicts. The fitted model reproduces the
median errors of the four flights at \rev{2.5~MHz and 5~MHz} within 25\%, and outlier rejection is the most effective
post-processing step, whereas filtering cannot remove the bias. Limitations include the cooperative transmitter, known altitude, one flight per configuration, unavailable TDOAs, and the model's underestimation of the heaviest error tails. Future
work will validate the model on more flights and sites, \rev{model gross errors and narrowband degradation}, and develop an estimator that
\rev{detects} and de-weights NLOS links.

\appendices

\section{Proof of Proposition~\ref{prop:ident}}\label{app:prop1}
The bound $\mathbf C_{\mathrm D}$ in~\eqref{eq:cD} is finite if and only if $\mathbf F_{\mathrm a}$
is invertible. Since $\mathbf Q_{\mathrm{noise}}^{-1}\succ0$, this holds if and only if
$\mathbf J_{\mathrm a}\in\Rt^{m\times(3+\KN)}$ has full column rank $3+\KN$, which requires
$m\ge3+\KN$, i.e.,~\eqref{eq:ident}. To show that the condition is tight and sufficient, we examine
the structure of $\mathbf J_{\mathrm a}$.

\emph{Case 1 (LOS reference):} By~\eqref{eq:dbias}, the column of $\mathbf J_b$ that corresponds to
the bias of an NLOS sensor $j\ne r$ is $-\mathbf e_\ell$, where $\mathbf e_\ell$ is the $\ell$th unit
vector and $\ell$ is the row with $i_\ell=j$. Ordering the rows so that the $m-\KN$ rows of the LOS
sensors come first, $\mathbf J_{\mathrm a}$ takes the block form
\begin{equation}
\mathbf J_{\mathrm a}=
\begin{bmatrix}
\mathbf J_{\mathcal L} & \mathbf 0\\
\mathbf J_{\mathcal N} & -\mathbf I_{\KN}
\end{bmatrix},
\label{eq:Jablock}
\end{equation}
where $\mathbf J_{\mathcal L}$ and $\mathbf J_{\mathcal N}$ contain the position rows~\eqref{eq:jac}
of the LOS and the NLOS sensors, respectively. Multiplying $\mathbf J_{\mathrm a}$ from the right by
the nonsingular matrix $\left[\begin{smallmatrix}\mathbf I_3 & \mathbf 0\\ \mathbf J_{\mathcal N} &
\mathbf I_{\KN}\end{smallmatrix}\right]$ yields the block-diagonal matrix with the blocks
$\mathbf J_{\mathcal L}$ and $-\mathbf I_{\KN}$, so
$\mathrm{rank}(\mathbf J_{\mathrm a})=\KN+\mathrm{rank}(\mathbf J_{\mathcal L})$. Full column rank
therefore holds if and only if $\mathrm{rank}(\mathbf J_{\mathcal L})=3$, which requires $m-\KN\ge3$,
i.e.,~\eqref{eq:ident}. For \rev{a non-degenerate geometry}, three LOS rows with linearly independent
vectors $\mathbf u_r-\mathbf u_{i_\ell}$ give rank three, which proves sufficiency.

\emph{Case 2 (NLOS reference):} If sensor $r$ is NLOS, the column of $b_r$ is the all-ones vector
$\mathbf 1$ by~\eqref{eq:dbias}, and the columns of the other $\KN-1$ NLOS biases are \rev{the negative unit vectors}
$-\mathbf e_\ell$ as in Case~1. Eliminating the NLOS rows with their own bias columns leaves the
$m-(\KN-1)=N-\KN$ rows of the LOS sensors, which determine the three coordinates and the offset
$b_r$. Full \rev{rank} then requires $\mathrm{rank}([\,\mathbf J_{\mathcal L}\ \ \mathbf 1\,])=4$
and hence $N-\KN\ge4$, again~\eqref{eq:ident}.

In both cases, each NLOS row carries its own free bias, which absorbs any value of that TDOA. The
row thus contributes nothing to the position, and the Schur complement of the bias block of
$\mathbf F_{\mathrm a}$ equals the FIM of the LOS rows alone, in agreement
with~\cite{qi2006analysis, shen2010fundamental}. When~\eqref{eq:ident} fails in Case~1,
$\mathbf J_{\mathcal L}$ has a nonzero null vector $\mathbf v$, and the joint perturbation
$(\delta\x,\delta\mathbf b)=(\mathbf v,\,\mathbf J_{\mathcal N}\mathbf v)$ satisfies
$\mathbf J_{\mathrm a}[\delta\x^{\mathsf T},\delta\mathbf b^{\mathsf T}]^{\mathsf T}=\mathbf 0$
by~\eqref{eq:Jablock}. Moving the position along $\mathbf v$ while changing the biases by
$\mathbf J_{\mathcal N}\mathbf v$ leaves every TDOA mean unchanged, so the two parameter values
cannot be distinguished, and no unbiased estimator has a finite variance. Case~2 follows in the same
way with a null vector of $[\,\mathbf J_{\mathcal L}\ \ \mathbf 1\,]$. Finally, when the altitude is
known, $\mathbf J$ has two columns, and the same argument with rank two in place of three gives
$N\ge3+\KN$. \hfill$\blacksquare$

\section{Proofs of Proposition~\ref{prop:floor} and Corollary~\ref{cor:bw}}\label{app:prop2}
\emph{Proof of Proposition~\ref{prop:floor}:} With the linearization
$\mathbf h(\tilde\x)\approx\mathbf h(\x)+\mathbf J(\tilde\x-\x)$, the WLS cost~\eqref{eq:wls} is
quadratic in $\tilde\x$. Setting its gradient to zero gives the normal equations
$\mathbf J^{\mathsf T}\mathbf W\mathbf J(\hat\x-\x)=\mathbf J^{\mathsf T}\mathbf W\big(\bm\rho-\mathbf h(\x)\big)$,
whose solution is
\begin{equation*}
\hat\x=\x+\mathbf G\big(\bm\rho-\mathbf h(\x)\big),\qquad
\mathbf G=\big(\mathbf J^{\mathsf T}\mathbf W\mathbf J\big)^{-1}\mathbf J^{\mathsf T}\mathbf W,
\end{equation*}
where $\mathbf J^{\mathsf T}\mathbf W\mathbf J$ is invertible for a non-degenerate geometry.
Inserting~\eqref{eq:rhomodel}, $\bm\rho-\mathbf h(\x)=\mathbf b_\rho+\mathbf n$, yields the error
\begin{equation*}
\hat\x-\x=\mathbf G\mathbf b_\rho+\mathbf G\mathbf n,
\end{equation*}
whose first term is deterministic and whose second term has zero mean. Taking the expectation
gives~\eqref{eq:pbias}, and the covariance of the estimate is
\begin{equation}
\mathrm{Cov}(\hat\x)=\mathbf G\,\mathbf Q_{\mathrm{noise}}\,\mathbf G^{\mathsf T}.
\label{eq:wlscov}
\end{equation}
Since the cross term $2\,\mathbf b_\rho^{\mathsf T}\mathbf G^{\mathsf T}\mathbf G\,\E[\mathbf n]$
vanishes, $\E\lVert\hat\x-\x\rVert^2=\lVert\mathbf G\mathbf b_\rho\rVert^2+\Tr(\mathbf
G\mathbf Q_{\mathrm{noise}}\mathbf G^{\mathsf T})$, which is the equality in~\eqref{eq:msefloor}.
For $\mathbf W=\mathbf Q_{\mathrm{noise}}^{-1}$, \eqref{eq:wlscov} reduces to
$(\mathbf J^{\mathsf T}\mathbf Q_{\mathrm{noise}}^{-1}\mathbf J)^{-1}=\mathbf C_{\mathrm n}$, and by
the Gauss--Markov theorem, no other weight matrix yields a smaller covariance, i.e.,
$\mathbf G\mathbf Q_{\mathrm{noise}}\mathbf G^{\mathsf T}\succeq\mathbf C_{\mathrm n}$. Taking the
trace gives the inequality in~\eqref{eq:msefloor}, and since the variance term is nonnegative,
$\mathrm{MSE}\ge\lVert\delta\x_{\mathrm b}\rVert^2$. \hfill$\blacksquare$

\emph{Proof of Corollary~\ref{cor:bw}:} By~\eqref{eq:snr}, $\mathrm{SNR}_i=P_{r,i}/(k_BTB)$, and
with $\beta=\kappa B$ for a constant $\kappa$, \eqref{eq:sigma} gives
$\sigma_i^2=c^2k_BT/(8\pi^2\kappa^2P_{r,i}B)$. Every ranging variance, and therefore every entry of
$\mathbf Q_{\mathrm{noise}}$, scales as $1/B$, which proves
$\mathbf Q_{\mathrm{noise}}(B)=(B_0/B)\,\mathbf Q_{\mathrm{noise}}(B_0)$. For the ML weight
$\mathbf W(B)=(B/B_0)\,\mathbf W(B_0)$, the factor $B/B_0$ appears once inside the inverse and once
outside it in $\mathbf G$, so it cancels, and $\delta\x_{\mathrm b}=\mathbf G\mathbf b_\rho$ does not
depend on $B$. In contrast, $\mathbf C_{\mathrm n}(B)=(B_0/B)\,\mathbf C_{\mathrm n}(B_0)$.
Substituting both into~\eqref{eq:msefloor} with equality gives~\eqref{eq:msebw}.
\hfill$\blacksquare$

\section*{Acknowledgment}
The authors used Anthropic's Claude Code and Cowork to assist with code development and simulations (Sections~\ref{sec:nlos}--\ref{sec:res}), data analysis, and drafting and editing of the manuscript text (all sections). All AI-assisted content was independently verified by the authors, who take full responsibility for the work. 

\bibliographystyle{IEEEtran}
\bibliography{references}

\end{document}